\documentclass[letterpaper,twocolumn,10pt]{article}

\usepackage{zhanggroup}
\newcommand{\mypara}[1]{\noindent{\bf {#1}.}\xspace}
\usepackage{hyperref}
\usepackage{tikz}
\usepackage{amsfonts}
\usepackage{amssymb}
\usepackage{pifont}
\usepackage{xspace} 
\usepackage{mathrsfs}
\usepackage{amsmath}
\usepackage{amsthm}
\usepackage{subcaption}
\usepackage{booktabs}
\usepackage{multirow}
\usepackage{graphicx} 
\usepackage[T1]{fontenc}
\usepackage{makecell}
\usepackage{bbm}
\usepackage[linesnumbered,ruled]{algorithm2e}
\begin{document}
% ----------------------------------------------------

\date{}

\title{Prompt Stealing Attacks Against Large Language Models}

\author{
Zeyang Sha\ \ \
Yang Zhang
\\
\\
\textit{CISPA Helmholtz Center for Information Security}
}

\maketitle

% ----------------------------------------------------
\begin{abstract}
% ----------------------------------------------------

The increasing reliance on large language models (LLMs) such as ChatGPT in various fields emphasizes the importance of ``prompt engineering,'' a technology to improve the quality of model outputs.
With companies investing significantly in expert prompt engineers and educational resources rising to meet market demand, designing high-quality prompts has become an intriguing challenge.
In this paper, we propose a novel attack against LLMs, named prompt stealing attacks.
Our proposed prompt stealing attack aims to steal these well-designed prompts based on the generated answers.
The prompt stealing attack contains two primary modules: the parameter extractor and the prompt reconstructor.
The goal of the parameter extractor is to figure out the properties of the original prompts.
We first observe that most prompts fall into one of three categories: direct prompt, role-based prompt, and in-context prompt.
Our parameter extractor first tries to distinguish the type of prompts based on the generated answers.
Then, it can further predict which role or how many contexts are used based on the types of prompts.
Following the parameter extractor, the prompt reconstructor can be used to reconstruct the original prompts based on the generated answers and the extracted features.
The final goal of the prompt reconstructor is to generate the reversed prompts, which are similar to the original prompts.
Our experimental results show the remarkable performance of our proposed attacks.
Our proposed attacks add a new dimension to the study of prompt engineering and call for more attention to the security issues on LLMs.

% ----------------------------------------------------
\end{abstract}
% ----------------------------------------------------

% ----------------------------------------------------
\section{Introduction}
% ----------------------------------------------------

With the rapid development of machine learning, large language models (LLMs) like GPT-4~\cite{O23} have risen to prominence as important tools in various fields, ranging from customer support~\cite{ZYJ23} and academic paper writing~\cite{GAG23, SWG23} to complex programming tasks~\cite{SBHP23,CMWWLGWYSXTD23}.
These models exhibit a remarkable ability to generate human-like text, providing immense value in diverse applications.
A growing number of individuals and organizations across different areas are beginning to take advantage of these LLMs in their daily work.
Nonetheless, the successful utilization of LLMs is not simply about deploying the models but entails a subtler, often overlooked aspect known as ``prompt engineering.''
The quality of output from these models heavily depends on the input queries or prompts provided to them.
Crafting high-quality prompts that can guide the model toward generating precise and meaningful responses is both an art and a science.

Recognizing the significance of high-quality prompts, an increasing number of companies are investing huge resources into hiring experts specialized in prompt engineering.
The expertise of these professionals lies in understanding the behavior pattern of LLMs and manipulating their inputs to achieve desired outputs.
For instance, company TikTok, among others, has established a dedicated team of prompt engineers, which has drastically improved the effectiveness of its AI-driven services.

Concurrent with this, there have been a lot of paid tutorials and courses aimed at educating individuals on the techniques of high-quality prompt design.
These educational resources underscore the growing interest and demand in the market for skills related to AI and machine learning, specifically in areas such as prompt engineering.
Several established prompt engineering techniques are currently in use, including direct prompts, where the model is directly instructed on what to do; in-context prompts, where additional information is provided to guide the response; and role-based prompts, where the prompts are designed as if the model is playing specific roles.
However, the application of these techniques and the creation of innovative prompts are heavily dependent on the nature of the tasks at hand.

\begin{figure*}[!t]
\centering
\includegraphics[width=2\columnwidth]{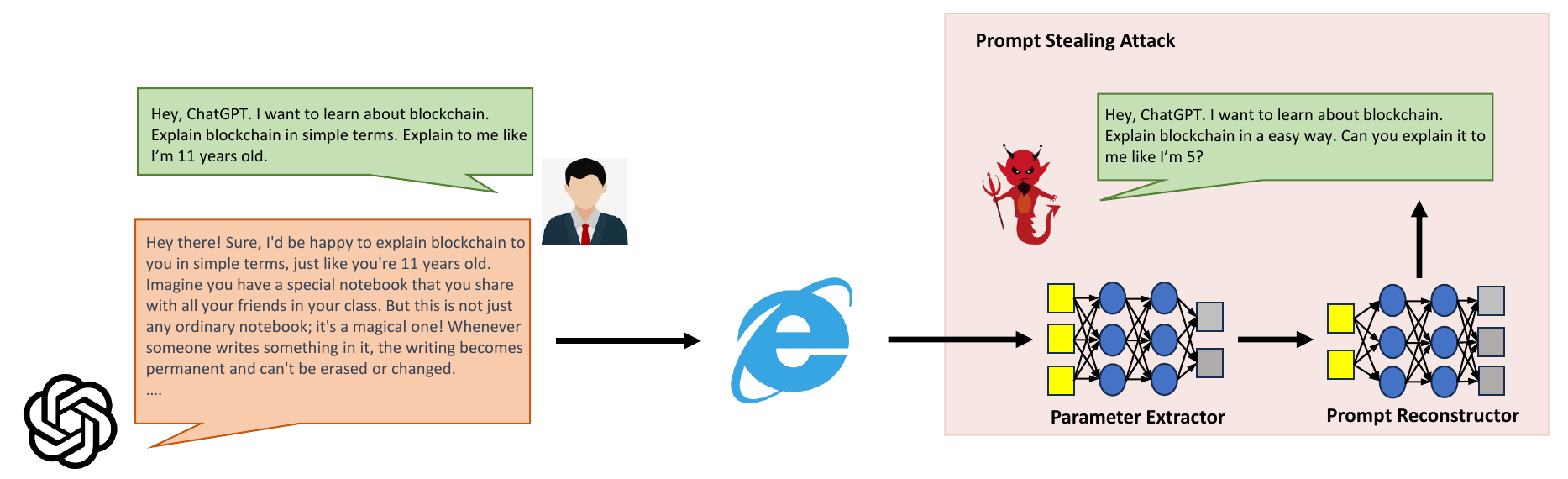}
\caption{The overview of the proposed prompt stealing attacks.
The users take advantage of prompt engineering to get  the desired answers from LLM.
Then, the adversary tries to reverse the original prompts through the parameter extractor and the prompt reconstructor.}
\label{fig:overview}
\end{figure*}

% ----------------------------------------------------
\subsection{Our Contribution}
% ----------------------------------------------------

As an increasing number of companies begin to make money by generating high-quality prompts for the users, a natural research problem emerged:  is it possible to extrapolate these prompts based on the responses generated by LLMs?
Viewing this from another perspective, such reverse-engineered prompts could be advantageous, serving as a stepping stone for auxiliary tasks like detecting fake text~\cite{MLKMF23,GZWJNDYW23,HSCBZ23} and generating jailbreak prompts~\cite{LGFXS23,DLLWZLWZL23,LDXLZZZZL23,SCBZ23}.
Therefore, in this work, we make the initial attempt to propose prompt stealing attacks against LLMs.

\mypara{Methodology}
Our proposed attack is shown in \autoref{fig:overview}.
As we have stated before, the goal of the prompt stealing attacks against LLMs is to reverse the original prompts according to the given answers.
To achieve this goal, we designed two modules that can work together: the parameter extractor and the prompt reconstructor.
The goal of the parameter extractor is to steal the detailed parameter information of the prompts according to the generated answer.
We first observe that most prompts fall into one of three categories: direct prompt, role-based prompt, and in-context prompt.
Direct prompt represents the simplest form where the user directly asks their desired questions.
Role-based prompt refers to the prompt that asks the LLMs to impersonate a specific role, such as when users use ChatGPT for movie reviews, starting prompts with ``Assume you are a movie commentator.''
In-context prompt involves providing certain contexts to help the LLM understand the desired topic.
Based on the above observation, we hope our parameter extractor can figure out which kind of prompts the original prompts are based on the generated answers.
Our parameter extractor first incorporates a three-class classifier named the primary classifier for the type determination.
Additionally, we construct two specialized sub-classifiers for the role-based prompts and the in-context prompts.
For role-based prompts, our sub-classifier is supposed to identify which role the LLM is pretending to play according to its generated answer.
We collect 15 of the most popular roles for the LLMs based on the questionnaires to cover the most common usage scenarios.
We then use ChatGPT to decide which different prompts are based on what roles and connect them as the new prompts.
The new prompts will be used to train our 15-class sub-classifier.
For the in-context prompts, we try to train another sub-classifier that can predict the number of contexts given by the original prompts.

Upon completing the parameter extraction stage, we have detailed insights into the original prompts.
We now try to reconstruct all the prompts based on the given answers and the features we obtained before the prompt reconstruction phase.
We first leverage ChatGPT to generate the directly reversed prompts based on the given answers.
Then, if the original prompts are predicted to be direct prompts according to the parameter extractor, we will just consider the output of ChatGPT to be the prompts we need.
If the original prompts are predicted to be role-based prompts, we will add the role prompt: ``Assume you are the predicted role,'' to the output of ChatGPT, and the added prompt can be considered as the new reversed prompt.
If the original prompts are predicted to be the in-context prompts, we will also leverage the ChatGPT again to generate a certain number of similar questions and answers according to the context number we have predicted.
Similar questions and answers together can be added to the output of ChatGPT as the new reversed prompts.

\mypara{Evaluation}
We perform experiments on two benchmark Question-Answering datasets, RetrievalQA~\cite{AAKV23} and Alpaca-GPT4~\cite{PLHGG23}, and two popular LLMs including ChatGPT~\cite{chatgpt} and LLaMA~\cite{TLIMLLRGHARJGL23}.

In the parameter extraction phase, our experimental results indicate that the proposed parameter extractor can achieve remarkable performance.
To be more specific, the primary classifier of the parameter extractor can achieve 0.833 accuracy to predict which type of prompts the original prompts are on RetrievalQA generated by ChatGPT.
Based on the results of the primary classifier, our two sub-classifiers for the role-based prompts and the in-context prompts can also achieve great performance.
For instance, the role-based sub-classifier can achieve 0.711 accuracy on the role-based prompts generated from the RetrievalQA datasets on LLaMA.
With the help of the primary classifier and the sub-classifiers, our parameter extractor can achieve overall excellent performance.
Note that to show the necessity of the primary classifier and the sub-classifiers, we consider the baseline model, which only takes the generated answer as the input and directly outputs the detailed information, i.e., how many contexts are used.
Our experimental results show that the current parameter extractor can achieve much better results than the baseline model, which indicates that hierarchical prediction is necessary.

In this prompt reconstruction phase, we consider two metrics to measure the attack's performance: prompt similarity and answer similarity.
Prompt similarity computes the cosine similarity between the reversed prompts and the original prompts.
Answer similarity refers to the similarity between the answers generated by the reversed prompts and the answers generated by the original prompts.
Note that we compute all the similarity based on the embedding generated by the sentence transformer~\cite{RG19}.
Our experimental results show that the proposed prompt reconstructor can have great performance.
For instance, on the prompts from RetrievalQA, the prompt reconstructor can achieve 0.832 prompt similarity on the answers generated by ChatGPT.
We also show that, armed with the information provided by the parameter extractor, the prompt reconstructor can achieve better performance.
For instance, if there is no parameter extractor, the prompt reconstructor can only achieve 0.616 answer similarity on the RetrievalQA datasets generated by ChatGPT for role-based prompts, but the similarity can be improved to 0.703 once we know the detailed parameter information.

We then make the first attempt to mitigate the vulnerabilities caused by the prompt stealing attacks proposed in this paper.
We consider two defense strategies.
The first defense needs the defender to add some prompts on the original prompts to prevent the generated answer containing the information from the prompts.
The second defense leverages ChatGPT to modify the generated answer to fool the prompt-stealing attacks with certain prompts.
Our extensive results indicate that both defenses can make the attack similarity drop.
However, compared to the baseline, the prompt similarity and the answer similarity are still high.
Therefore, we can argue that the proposed prompt stealing attack has a certain level of robustness.

\mypara{Implications}
In this paper, we propose the first prompt stealing attacks against language learning models (LLMs).
Our work uncovers the risks associated with these attacks on widely used LLMs.
We believe that the research findings presented here can provide valuable guidance for developing robust LLM systems.
Furthermore, we aim to heighten attention and increase awareness about the safety and security issues inherent to LLMs.

% ----------------------------------------------------
\section{Preliminary}
% ----------------------------------------------------

% ----------------------------------------------------
\subsection{Large Language Models}
% ----------------------------------------------------

LLMs are generation models that can generate texts conditioned by giving them the designed prompts.
These kinds of models are trained on a large number of online texts with unsupervised or reinforcement learning.
The proposal of transformer~\cite{VSPUJGKP17} can be regarded as the starting point of the advancement of LLMs.
The key idea behind the transformer is the self-attention mechanisms, which allow the models to focus on important parts of the input instead of the entire text.
Based on transformers, a generative pre-trained model (GPT)~\cite{AKTI20} was proposed to deal with different downstream tasks in the NLP domain.
It takes advantage of the pre-training and then fine-tuning module for different tasks.
GPT achieves great performance in a wide range of fields.
Based on the structure of GPT, GPT2~\cite{RWCLAS19} and GPT3~\cite{BMRSKDNSSAAHKHCRZWWHCSLGCCBMRSA20} were proposed one after another with larger parameters and larger training datasets.
These models have shown an excellent ability to deal with tasks in different fields.
At the same time, they also initially showed that they have a certain ability for independent inference and thinking.
What makes LLMs really popular is the proposal of ChatGPT in November 2022.
It is based on the GPT-3.5 architecture and optimized with Reinforcement Learning from Human Feedback (RLHF)~\cite{CLBMLA17,SOWZLVRAC20}
to make the model truly intelligent.
ChatGPT~\cite{chatgpt} shows revolutionary capabilities when facing almost all daily life scenarios.

% ----------------------------------------------------
\subsection{Prompt Engineering}
% ----------------------------------------------------

Prompt engineering is a newly existing technology for developing and optimizing prompts to better leverage LLMs for users' specific tasks.
Typically, prompts can be classified as direct prompts, role-based prompts, and in-context prompts.
For the role-based prompts, previous work has shown that with the proper role, LLMs can be used to generate toxic contents~\cite{DFWUKW20, DABSHBR21, SBBCSZZ22}, game designing~\cite{POCMLB23}, and so on.
Also, most jail-break prompts are also role-based prompts~\cite{KLSGZH23}.
For the in-context prompts, previous works have found that LLMs have the ability to conduct few-shot learning.
There are many new proposed in-context learning technology to boost the performance of LLMs like chain-of-thought context~\cite{WWSBIXCLZ22,WWSLCNCZ23} and tree-of-thought context~\cite{ZWMG22,YYZSGCN23}.
However, since there is no consensus on which type of context is better, in this work, we only consider the most obvious in-context prompts: directly using the text as the context.

% ----------------------------------------------------
\section{Threat Model}
% ----------------------------------------------------

In this paper, we refer to the prompts that were used to generate the given answer as original prompts and the prompts we have stolen as reversed prompts.
We will first introduce the adversary's goal and then emphasize the background knowledge the adversary can obtain.

% ----------------------------------------------------
\subsection{Adversary's Goal}
% ----------------------------------------------------

To successfully steal the original prompts based on the given answer, the goal of prompt stealing attacks should follow the following points.

\begin{itemize}
\item \mypara{Exact Parameter Information}
The first goal of the adversary is to try to infer the exact parameter information according to the generated answers.
To be more specific, the adversary should be able to predict which kind of prompts the original prompts are.
The knowledge about parameter information can foster the whole process of prompt stealing.
\item \mypara{Similar Reversed Prompt}
The reversed prompts should be similar to the original prompts.
They should possess similar semantics and share similar structures.
Also, the reversed prompt should be able to generate a similar answer as the original prompts.
This is also the primary goal of our proposed prompt stealing attacks.
\end{itemize}

% ----------------------------------------------------
\subsection{Adversary's Background Knowledge}
% ----------------------------------------------------

We assume the adversary has the following background knowledge regarding the original prompts.

\begin{itemize}
\item \mypara{Access to the Generated Answers}
We assume the adversary can have access to the answers generated by the original prompts.
Note that the adversary should have no knowledge about the original prompts.
\item \mypara{Knowledge About LLMs}
Regarding the knowledge of LLMs that are leveraged to generate answers, we assume the adversary can know what kind of LLMs are used.
It is a reasonable assumption as most of the answers shared on the internet have classified which model they are used to generate.
\end{itemize}

% ----------------------------------------------------
\section{Prompt Stealing Attack}
% ----------------------------------------------------

In this section, we introduce the proposed prompt stealing attacks against large language models.
We first describe how we steal the detailed parameter information through the proposed parameter extractor module.
Then, we further introduce the prompt reconstructor module, where we can reverse the original prompts.

% ----------------------------------------------------
\subsection{Parameter Extraction}
% ----------------------------------------------------

\begin{figure}[!t]
\centering
\includegraphics[width=1\columnwidth]{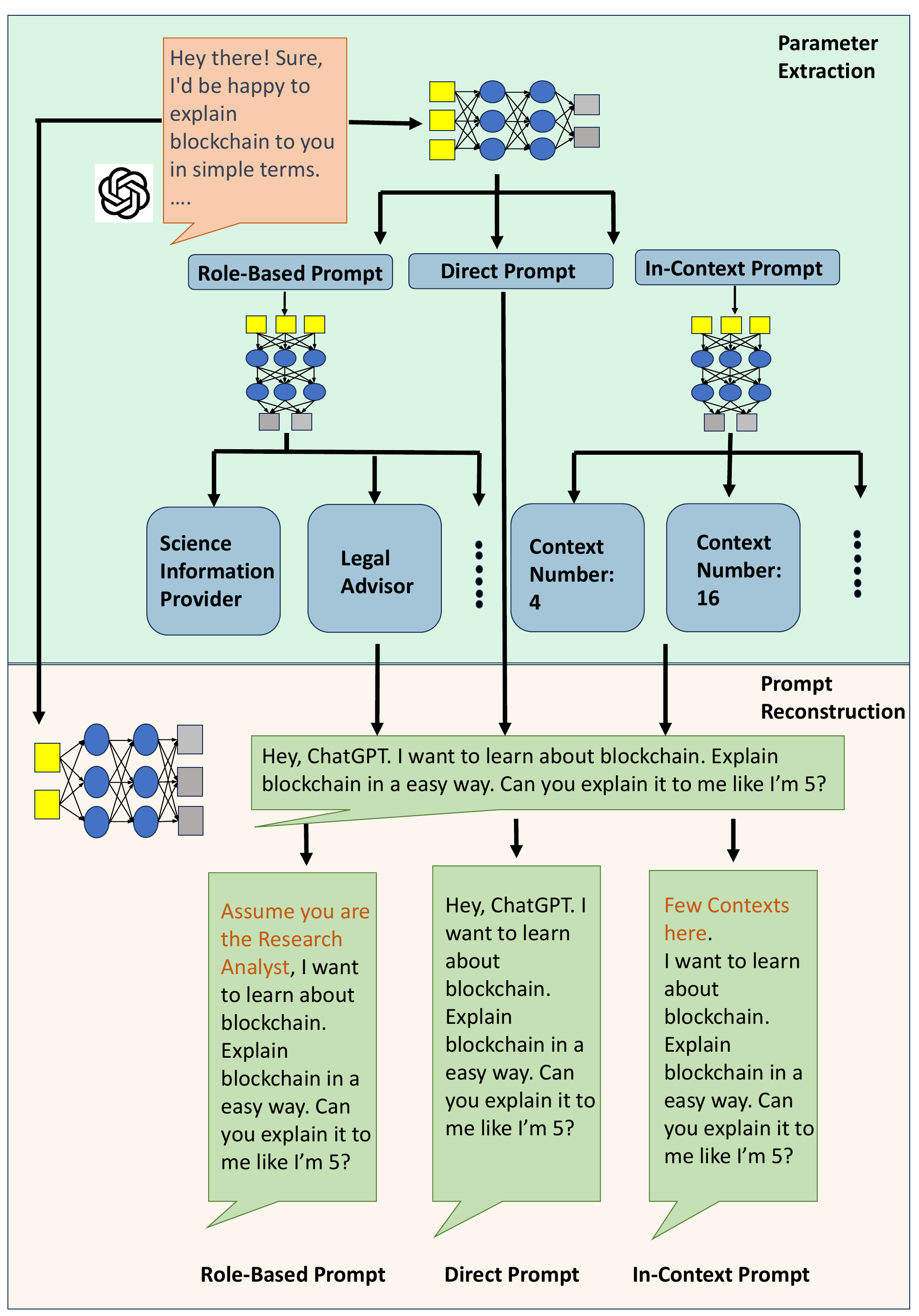}
\caption{The overview of the method we have proposed.
It involves two stages: parameter extraction and prompt reconstruction.
In the parameter extraction phase, the extractor tries to predict the type of prompts and further information.
With the parameter information and the given answers, the prompt reconstructor tries to generate the reversed prompts.}
\label{fig:method}
\end{figure}

\mypara{Overview}
The basic idea of parameter extraction is to infer the parameters of the prompts according to the generated answers.
The detailed parameters of the prompt can be defined as the type and the further structure information of the original prompt.
To effectively implement the functionality of parameter extraction, we first conduct the prompt categorization through the questionnaire answered by 10 users who use ChatGPT frequently.
Based on the results of our questionnaire, we find that most popular prompts fall into three categories, including direct prompts, role-based prompts, and in-context prompts.
Then, we construct the training dataset containing different kinds of prompts to train our classifiers.
Our parameter extractor contains three classifiers: one primary classifier and two sub-classifiers.
The primary classifier aims to determine which type of prompt is.
The sub-classifiers aim to determine further information based on the prediction of the primary classifier.
In the end, the proposed parameter extraction can be used to predict which kinds of prompts the original prompts are and the detailed structure information.

\begin{table*}[!t]
\centering
\caption{Different types of prompts.}
\label{table:prompttype}
\scalebox{0.85}{
\begin{tabular}{l|p{.3\linewidth}|p{.3\linewidth}}
\toprule
\textbf{Prompt Type} & \textbf{Description} & \textbf{Example} \\
\midrule
Direct Prompts &  Direct prompts are straightforward prompts that directly ask questions.
It typically requires LLMs to answer or complete a specific task without any additional context. & Why do humans have different colored eyes? 
What causes people to have different colors of eyes? 
Is there a point to the color of your eyes other than superstitions? \\
\midrule
Role-Based Prompts  &  Role-based prompts assign specific roles to the LLMs to help the models fully understand the desired questions.
It is one of the most commonly used categories of prompts. & I want you to act as an etymologist. 
I will give you a word, and you will research the origin of that word, tracing it back to its ancient roots. 
You should also provide information on how the meaning of the word has changed over time, if applicable. 
My first request is "I want to trace the origins of the word 'pizza.'" \\
\midrule
In-Context Prompts &   In-context prompts incorporate a few demonstration examples that are similar to the desired task in the prompts based on the principle of few-shot learning. &  The odd numbers in this group add up to an even number: 17,  9, 10, 12, 13, 4, 2.
A: Adding all the odd numbers (17, 9, 13) gives 39. The answer is False. \
The odd numbers in this group add up to an even number: 15, 32, 5, 13, 82, 7, 1. 
A: \\
\bottomrule
\end{tabular}
}
\end{table*}

\mypara{Prompts Categorization}
We first categorize prompts based on their types and structures.
We designed the questionnaire to ask 10 users who use ChatGPT frequently about what types of prompts they always use.
Then, we collect the answers and summarize the results.
According to the results of the questionnaire, we find that the most popular prompts fall into three main types: direct prompts, role-based prompts, and in-context prompts.
We show some examples in \autoref{table:prompttype}.
Based on the categorization, we will next construct the datasets and the parameter extractor.

\mypara{Data Collection}
We then try to collect different prompts to construct our dataset.
In order to align different prompts in their distribution, we leverage one direct prompt dataset and expand it to the role-based prompt and in-context prompt.
We take the following steps for the expansion:

\begin{itemize}
\item \mypara{Role-Based Prompts}
We first collect 15 roles that are most commonly used recently.
It is collected by the questionnaire answered by 10 users who use ChatGPT frequently.
It is concluded in \autoref{table:commonroles} in the Appendix.
Regarding how to assign different roles to different prompts, we consider two cases.
In the first case, we assign the roles to different prompts according to their relevance according to the rating of ChatGPT.
In this case, we design the prompt: ``Which following role [listed role] are you assumed to be.''
This is the most common case due to that the role-based prompt tends to find the most suitable role.
In the second case, we assign the collected roles at random to observe how the role affects the generated answer.
\item \mypara{In-Context Prompts}
To generate the in-context prompts, we randomly select a certain number of existing prompts and corresponding answers as the context and add them to the original prompts.\
We try to figure out whether different contexts have impacts on the generated answer.
\end{itemize}

\mypara{Parameter Extraction}
We then construct our parameter extractor according to the above categorization and data collection as shown in \autoref{fig:method}.
Our parameter extractor consists of one primary classifier and two sub-classifiers.
The primary classifier aims to predict which kind of prompt the original prompt is.
We build the primary classifier (i.e., a three-class classifier) that accepts the generated answers as the input and outputs the predicted type, i.e., direct prompt.
We take advantage of the pre-trained BERT~\cite{DCLT19} model as the backbone of our primary stealer.
Note that to evaluate the generalizability of the trained primary classifier, we test the primary classifier on unknown prompt datasets.
Based on the prediction result of the primary classifier, we train two sub-classifiers for the detailed parameter information.
If the prompt is predicted to be a role-based prompt, one sub-classifier will be leveraged to determine which role the prompt belongs to.
This sub-classifier is trained on the role-based data we built before.
If the prompt is predicted to be the in-context prompt, another sub-classifier can be used to further tell how many contexts are used in the original prompts.
This sub-classifier is trained on the in-context data mentioned before.
In general, the proposed parameter extractor can be used to infer the parameter information of the original prompts through the well-constructed primary classifier and two sub-classifiers.

% ----------------------------------------------------
\subsection{Prompt Reconstruction}
% ----------------------------------------------------

\mypara{Overview}
In the prompt reconstruction phase, we aim to reconstruct the whole prompt, which should be similar to the original prompts.
We take advantage of the previous classifier's results by designing a prompt reconstruction module.
We first leverage ChatGPT to directly generate the reversed prompts according to the given answers.
Then, we add different information based on the output of the previous parameter extractor.
We will show later in the experiment phase that the proposed prompt reconstruction process can achieve great performance.

\mypara{Direct Prompt Reconstruction}
We leverage ChatGPT To generate direct prompts based on the given answers.
We design the reconstruction prompts: ``What question are you asked if you can generate the following answer?''
Then we input the given answer, and the ChatGPT will return the directly reversed prompt.

\mypara{Assembly Composition}
After the generation, we can construct the reversed prompt based on the detailed parameter information we obtained before.
To be more specific, we take the following steps for different types of prompts:

\begin{itemize}
\item \mypara{Role-Based Prompt}
If the original prompt is the role-based prompt, we will add ``Assume you are a '' together with the predicted role to the directly reversed prompt to get a new role-based reversed prompt.
\item \mypara{In-Context Prompt}
If the original prompt is the in-context prompt, we will leverage ChatGPT to generate other contexts based on the number of previously predicted results.
Note that the prompt to generate context can be designed as ``Can you generate [predicted number] questions and corresponding answers similar to [the directly reversed prompt].''
The generated context, together with the directly reversed prompt, can be considered as the new reversed prompt.
\end{itemize}

% ----------------------------------------------------
\subsection{Summary}
% ----------------------------------------------------

In conclusion, our proposed prompt stealing attacks against LLMs are composed of the parameter extractor and the prompt reconstructor.
The goal of our parameter extraction is to predict which type of prompts is used to generate the given answers.
We construct our parameter extractor using one primary classifier and two sub-classifiers.
The primary classifier can be used to predict whether the original prompts belong to the direct prompt, role-based prompt, or in-context prompt.
Then, based on the output of the primary classifier, we further develop two sub-classifiers to further determine which role is used if the original prompt is the role-based prompt and how many contexts are used if the original prompt is the in-context prompt.
Together, the combination of the primary classifier and the sub-classifiers can be considered our parameter extractor.
The goal of the prompt reconstructor is to reverse the exact prompt that should be similar to the original prompts.
We take advantage of the information provided by the parameter extractor and leverage ChatGPT to generate the prompts.
We show in further experiments that the proposed parameter extractor and prompt reconstructor can have much better performance than the baseline.

% ----------------------------------------------------
\section{Experiments}
% ----------------------------------------------------

In this section, we first introduce the experimental setup for the proposed attacks, including the target models, datasets, and evaluation metrics.
Then, we introduce the performance of the parameter extractor and the prompt reconstruction.

% ----------------------------------------------------
\subsection{Experimental Setup}
% ----------------------------------------------------

\mypara{Large Language Models}
To evaluate the effectiveness of the proposed prompt stealing attacks, we leverage two different kinds of large language models, including ChatGPT and LLaMA, for the experiments.

\begin{itemize}
\item \mypara{ChatGPT~\cite{chatgpt}}
ChatGPT is the advanced large language model based on GPT-3.5's architecture.
It can generate human-like answers for different tasks.
ChatGPT is optimized by the Reinforcement Learning from Human Feedback (RLHF) to achieve great performance.
\item \mypara{LLaMA~\cite{TLIMLLRGHARJGL23}}
LLaMA is another state-of-the-art large language model released by Meta.
It is designed to generate human-like texts, solve mathematical theorems, and other tasks.
LLaMA is also based on the transformer and can be considered as the next word predictor.
Different from ChatGPT, LLaMA is open source and can be deployed on a personal server.
\end{itemize}

Note that all of our models will only be trained on the ChatGPT's data.
LLaMA will be used to test the generalizability of the proposed attacks, as it can serve as a different target model.

\mypara{Prompt Datasets}
Currently, there are numerous studies that aim to evaluate the performance of LLMs.
Therefore, in this work, we take advantage of existing question-answering datasets to conduct our attack.

\begin{itemize}
\item \mypara{Alpaca-GPT4~\cite{PLHGG23}}
The Alpaca-GPT4 was introduced to replicate the performance of ChatGPT and GPT-4.
It covers a wide range of usage scenarios for LLMs.
Therefore, Alpaca-GPT4 is used in this work as the original prompts datasets.
\item \mypara{RetrievalQA~\cite{AAKV23}}
RetrievalQA is another dataset built for the evaluation of using generative LLMs to augment training data for retrieval models in different domains.
It contains question prompts from different aspects of daily life.
\end{itemize}

Note that due to the fact that querying ChatGPT costs lots of time and money, we only randomly selected 600 prompts from the two datasets.
For the RetrievalQA, we divide 500 prompts as the training prompts for our attack.
The remaining prompts from RetrievalQA and all prompts we have selected from Alpaca-GPT4 are considered test prompts.

\mypara{Evaluation Metric}
For the parameter extraction, we use accuracy, recall, precision, and AUC to evaluate different prediction tasks.
For the prompt reconstruction, we leverage the following metrics for the evaluation.

\begin{itemize}
\item \mypara{Prompts Similarity (PS)}
As our primary goal is to make the reversed prompts similar to the original prompts, computing the similarity between the two prompts is the most obvious metric.
We leverage the sentence transformer to generate the embeddings for both reversed and original prompts and then compute the similarities between them.
More specifically, we compute the similarities as follows:
\[ \text{cosine\_similarity}(\mathbf{R_P}, \mathbf{O_P}) = \frac{\mathbf{R_P} \cdot \mathbf{O_P}}{\|\mathbf{R_P}\| \cdot \|\mathbf{O_P}\|}
\]
where $R\_P$ represents reversed prompts and $O\_P$ represents original prompts.
\item \mypara{Answers Similarity (AS)}
Note that the motivation for the prompt stealing is to reverse the prompt, which can have similar answers to the original prompts.
Therefore, evaluating the similarities between the generated answers of reversed prompts and original prompts can also be considered an effective metric.
To compute such similarity, we first generate two answers based on the reversed and original prompts.
Then, similar to the above answer similarity computing process, we leverage a sentence transformer to generate two embeddings and compute the similarities as follows:
\[ \text{cosine\_similarity}(\mathbf{R_A}, \mathbf{O_A}) = \frac{\mathbf{R_A} \cdot \mathbf{O_A}}{\|\mathbf{R_A}\| \cdot \|\mathbf{O_A}\|}
\]
where R\_A represents answers generated by reversed prompts and O\_A represents answers generated by original prompts.
\end{itemize}

\begin{table*}[!t]
\centering
\caption{Performance of the parameter extraction.}
\label{table:performance-promptsteal}
\scalebox{0.85}{
\begin{tabular}{l|ccccccc}
\toprule
\textbf{Defense Method} & \textbf{Models} & \textbf{Datasets} & \textbf{Accuracy} & \textbf{Precision} & \textbf{Recall} & \textbf{AUC} & \textbf{Random Guess} \\
\midrule
\multirow{4}*{Primary Classifier} & ChatGPT & RetrievalQA & 0.833 & 0.845 & 0.798 & 0.832 & 0.333 \\
~  & ChatGPT & Aplaca-GPT4 & 0.811 & 0.825 & 0.773 & 0.791 & 0.333 \\
~ & LLaMA & RetrievalQA & 0.884 & 0.912 & 0.899 & 0.901 & 0.500 \\
 ~ & LLaMA & Aplaca-GPT4 & 0.855 & 0.864 & 0.841 & 0.888 & 0.500 \\
 \midrule
\multirow{4}*{Sub-classifier for the role-based prompts} & ChatGPT & RetrievalQA & 0.732 & 0.743 & 0.728 & 0.711 & 0.067 \\
~  & ChatGPT & Aplaca-GPT4 & 0.695 & 0.701 & 0.689 & 0.742 & 0.067 \\
~ & LLaMA & RetrievalQA & 0.711 & 0.734 & 0.703 & 0.695 & 0.067 \\
 ~ & LLaMA & Aplaca-GPT4 & 0.673 & 0.634 & 0.713 & 0.699 & 0.067 \\
 \midrule
 \multirow{2}*{Sub-classifier for the in-context prompts} & ChatGPT & RetrievalQA & 0.614 & 0.643 & 0.581 & 0.622 & 0.250 \\
~  & ChatGPT & Aplaca-GPT4 & 0.601 & 0.624 & 0.589 & 0.611 & 0.250 \\
\bottomrule
\end{tabular}
}
\end{table*}

% ----------------------------------------------------
\subsection{Performance of Parameter Extraction}
% ----------------------------------------------------

In this section, we first show the performance of the primary classifier for parameter extraction.
Then, we evaluate the proposed two sub-classifiers based on the prediction results from the primary classifier.

\mypara{Evaluation Results for Primary Classifier}
We show the evaluation result for the primary classifier in \autoref{table:performance-promptsteal}.
the goal of the primary classifier is to determine whether the given answer is generated by the direct prompt, role-based prompt, or in-context prompt.
Note that due to the limitation of our computing resources and the input limitation of LLaMA, the LLaMA model cannot take the long context as the input.
Therefore, in the results shown in \autoref{table:performance-promptsteal}, the primary classifier for LLaMA can be considered as the two-class classifier's results which only contain direct prompts and role-based prompts.

From the table, it can be seen that the proposed classifier can achieve great performance in all the cases.
For instance, to determine the prompts of the answers generated by ChatGPT, the classifier can achieve 0.833 accuracy on RetrievalQA datasets and 0.811 performance on Alpaca-GPT4 datasets.
We emphasize here that for different models (ChatGPT and LLaMA), we train the classifier on one dataset (RetrievalQA) and then test it on both the RetrievalQA test set and Alpaca-GPT4.
The performance on Alpaca-GPT4 further demonstrates that the trained primary classifier shows great generalization ability against the different distributions of original prompts.
Also, we can find that, overall, the performance of the primary classifier (3-class classification) on ChatGPT is similar to the performance on LLaMA, while the primary classifier for LLaMA is the 2-class classification.
That is because, compared to ChatGPT, the output generated by LLaMA is not that clear and related to the original prompts.
We show one example in \autoref{table:diff_chat_llama}.
Therefore, the difference between the performance of the proposed attacks on the two models can be considered as the difference between the generation abilities of the two models.
It can be concluded that the proposed primary classifier can work very well on the two current popular large language models.

\begin{table*}[!t]
\centering
\caption{Difference between ChatGPT and LLaMA.}
\label{table:diff_chat_llama}
\scalebox{0.85}{
\begin{tabular}{p{.3\linewidth}|p{.3\linewidth}|p{.3\linewidth}}
\toprule
\textbf{Prompts} & \textbf{ChatGPT Answer} & \textbf{LLaMA Answer} \\
\midrule
Are there any good museums in Istanbul for me to visit? & Istanbul is rich in history and culture, offering a wide range of fantastic museums. Here are some of the top ones you can consider: (1) Pera Museum (2) Istanbul Modern ... & 1. Archeology Museum: A must-see. 2. Topkapi Palace: A must-see. 3. Aynalikavak Palace: A must-see. 4. Rustem Pasha Mosque: A must-see. \\
\bottomrule
\end{tabular}
}
\end{table*}

\mypara{Evaluation Results for Sub-Classifiers}
Based on the prediction results, we train our sub-classifiers for different downstream tasks.
We now show the performance of our sub-classifiers in this section.
Our proposed sub-classifiers have different functions for different prediction results.
For instance, if the primary classifier predicts that the given answers are generated by the role-based prompt, then one sub-classifier can be used to predict which role is used in the original prompts.
If the prediction results of the primary classifier are the in-context prompts, then another sub-classifier should be leveraged to predict the context number used in the original prompts.
Otherwise, if the original prompts are just direct prompts, then no sub-classifier is needed for further prediction.
\autoref{table:performance-promptsteal} can summarize the performance of sub-classifiers for different tasks.

It can be seen from the table that our sub-classifiers for different tasks can still achieve outstanding performance.
For instance, to predict which role is used in the original prompts, the sub-classifier can achieve 0.732 accuracy on answers from RetrievalQA generated by ChatGPT, which is a 15-class classification task.
From the results, we can also conclude that the sub-classifier for the context number has worse performance than the sub-classifiers for the role.
For instance, as the 13-class classification, the sub-classifier for the role can still have better performance (0.732) than the sub-classifier for context number (0.614), which is the 4-class classification.
The results reveal that the set role has much more impact on the quality of the generated answers than the number of contexts.
However, whether the prompts contain the context or not can still have a significant impact on the generated answers because the primary classifier can achieve great performance.
Also, as LLaMA cannot take inputs that are too large, we only test the direct prompt and role-based prompt, as we stated before.

\begin{figure}[!t]
\centering
\begin{subfigure}{0.49\columnwidth}
\includegraphics[width=\columnwidth]{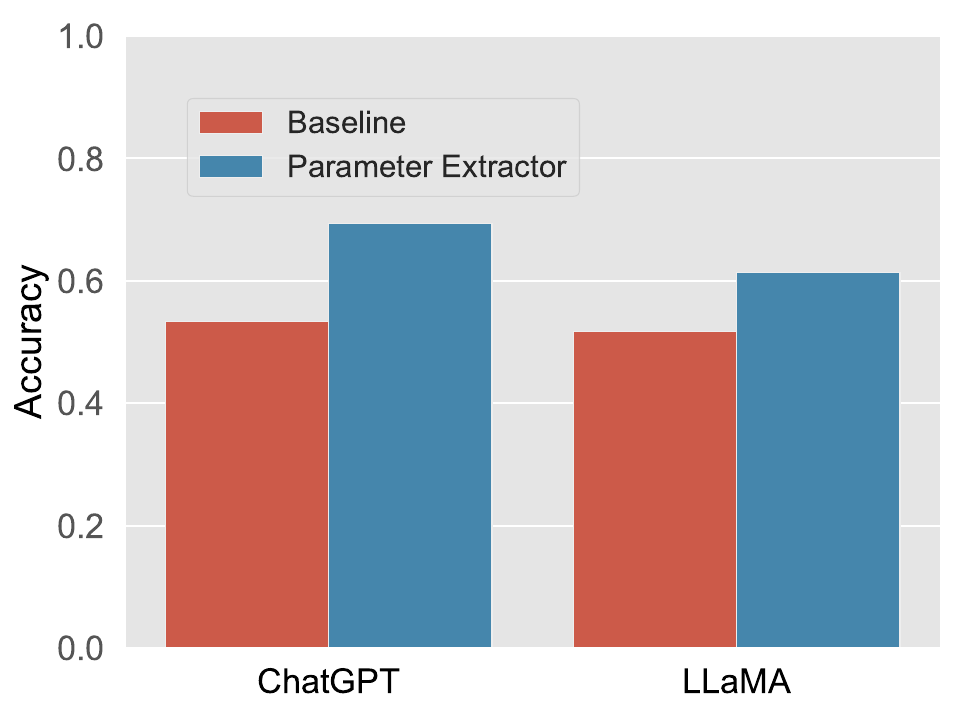}
\caption{RetrievalQA}
\label{fig:ChatGPT+RetrievalQA}
\end{subfigure}
\begin{subfigure}{0.49\columnwidth}
\includegraphics[width=\columnwidth]{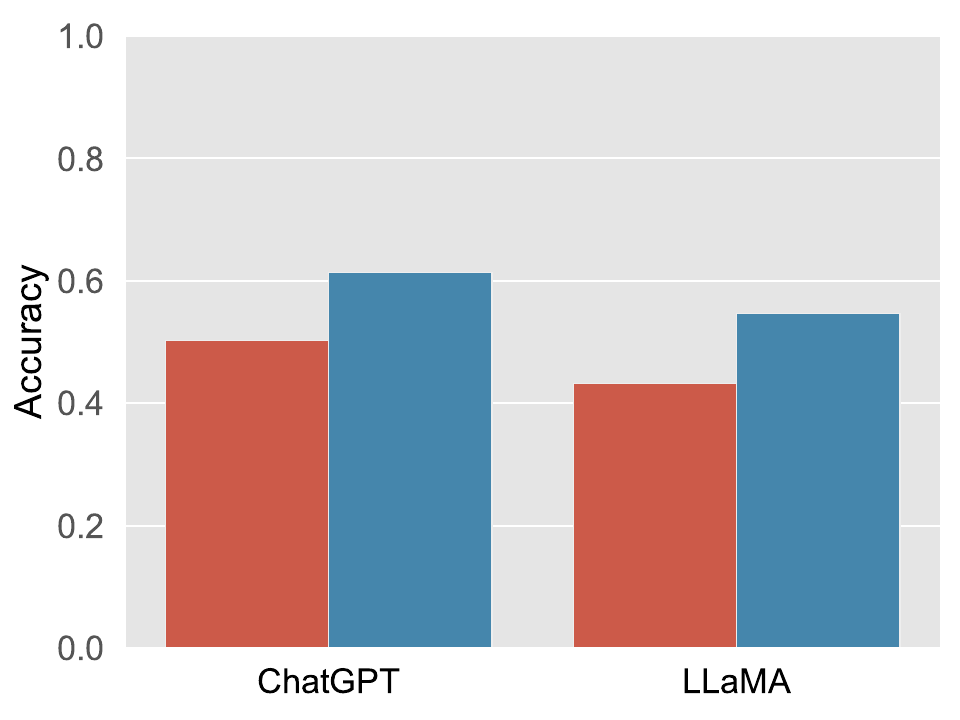}
\caption{Alpaca-GPT4}
\label{fig:ChatGPT+Alpaca-GPT4}
\end{subfigure}
\caption{Overall performance of the parameter extractor.}
\label{fig:overallperformance}
\end{figure}

\mypara{Overall Performance}
Together with the primary classifier and the sub-classifiers, we construct our parameter extractor.
The above section shows the outstanding performance of different models used in the parameter extractor.
In this section, we show the overall performance of the whole module.

The results are summarized in \autoref{fig:overallperformance}.
It can be seen that with different classifiers together, our proposed parameter extractor can have great performance.
Note that we show the baseline results by directly training the 20-class classifier as the prompt extractor.
The baseline model can directly predict which role or how many contexts are contained in the original prompts in one model as the 20 classification tasks.
We can see from the figure that the idea of using different classifiers to conduct the hierarchical prediction can largely boost the performance of the parameter extractor.
For instance, the overall performance on RetrievalQA generated by ChatGPT can achieve 0.693 accuracy, while the baseline can only achieve 0.234 accuracy.
Note that the hierarchical parameter extractor excels not just in terms of performance but also has extra advantages when utilized with new kinds of prompts associated with frequently employed roles.

\mypara{Takeaways}
In conclusion, the proposed parameter extractor can achieve great performance on the existing popular LLMs and different distributed datasets.
To be more specific, as the first part of our extractor, the primary classifier aims to predict which type of original prompts can achieve outstanding performance even on the different distributed data.
The following sub-classifiers can also work very well in their different tasks.
Together with the primary classifier and the sub-classifiers, our parameter extractor can work very well to steal the parameter information of the original prompts based on the generated answers.

\begin{table*}[!t]
\centering
\caption{The performance of parameter reconstruction.}
\label{table:para-reconstruct}
\scalebox{0.85}{
\begin{tabular}{lcccccc}
\toprule
~ & LLM & Types of the prompt & RetrievalQA & Baseline & Alpaca-GPT4 & Baseline  \\
\midrule
\multirow{5}*{Prompt Similarity} & \multirow{3}*{ChatGPT} & Direct Prompt & 0.832 & 0.832 & 0.817 & 0.817 \\
~  & ~ & Role-Based Prompt & 0.803 & 0.785 & 0.799 & 0.792 \\
~  & ~ & In-Context Prompt & 0.501 & 0.450 & 0.562 & 0.489\\
~  & \multirow{2}*{LLLaMA} & Direct Prompt & 0.794 & 0.794 &  0.712 & 0.712 \\
~  & ~ & Role-Based Prompt & 0.745 & 0.802 & 0.773 & 0.711 \\
\midrule
\multirow{5}*{Answer Similarity} & \multirow{3}*{ChatGPT} & Direct Prompt & 0.768 & 0.768 & 0.754 & 0.754 \\
~  & ~ & Role-Based Prompt & 0.703 & 0.616 & 0.699 & 0.658 \\
~  & ~ & In-Context Prompt & 0.523 & 0.712 & 0.562 & 0.615\\
~  & \multirow{2}*{LLLaMA} & Direct Prompt & 0.706 & 0.706 & 0.712 & 0.712 \\
~  & ~ & Role-Based Prompt & 0.756 & 0.743 & 0.732 & 0.695 \\
\bottomrule
\end{tabular}
}
\end{table*}

% ----------------------------------------------------
\subsection{Performance of Prompt Reconstruction}
% ----------------------------------------------------

We show the great performance of the proposed parameter extractor in the previous section.
In this section, based on the results of the parameter extractor, we further show the performance of the proposed prompt reconstructor.
Our proposed prompt reconstructor aims to reverse the exact prompts that should be similar to the original prompts.
We leverage prompts similarity (PS) and the generated answer similarity (AS) as the metrics to conduct the evaluation.

We show the results in \autoref{table:para-reconstruct}.
It can be seen that apart from the sentence similarity, the structure and the tone used in the original sentence are also perfectly imitated.
As we take advantage of ChatGPT for the reconstruction task, it can be concluded that ChatGPT has a remarkable ability in reverse engineering.
For instance, when using ChatGPT to generate the reversed prompt for the direct prompt, the model can achieve 0.832 PS and 0.768 AS on ChatGPT.
Although ChatGPT is good at prompt reconstruction, we can also conclude from the results that the information provided by the parameter extractor can boost the performance of prompt reconstruction.
To show the necessity of the previous parameter extractor, we consider the naive version of the prompt stealing attack as the baseline for the prompt reconstruction, which just uses the generated answers to generate the reversed prompts.
Note that as the baseline is just the case for the direct prompt, the results of the baseline and the direct prompt are the same.
It can be seen from the table that the proposed prompt reconstruction can achieve great performance with high PS and high AS.
For instance, the reconstructor can achieve 0.803 PS and 0.703 AS when reversing the role-based prompt from RetrievalQA generated by ChatGPT.
Hence, we can postulate that together with the ChatGPT's remarkable reversing ability and the information we obtained from the parameter extractor, we can conduct very successful prompt stealing attacks against LLMs.

Note that as we can not reverse the exact context for the in-context prompt, the PS for the in-context prompt is very low.
This is due to the fact that the answer response to the in-context prompt cannot reveal the context information that is randomly selected from the original datasets.
The role of the context is to help the model better understand and grasp the train of thought for problem-solving.
Therefore, although the PS is very low, the AS for the in-context prompt is good enough to be considered a successful attack.

Different from the in-context prompt, the reversed role-based prompt can achieve better performance in both PS and AS.
This result indicates that for the role-based prompt if the reversed prompt is more similar to the original prompt, the generated answer can be more similar to the original answer.
If we take into account all the results collectively, it can be concluded that the AS is the better metric for prompt reconstruction tasks.
This is due to the fact that the final goal of the prompt stealing attacks is to generate prompts that can have similar corresponding answers as the original prompts.
Therefore, even if the PS is not that high (i.e., reversed in-context prompt), the similar generated answer can still be regarded as a successful attack.

\mypara{Takeaways}
In conclusion, we perform the evaluation on the prompt reconstruction.
Our results show that the proposed prompt reconstructor can have great performance in reversing the exact prompts.
Also, we find that AS can be considered a better metric to measure the quality of the reversed prompts than the prompt similarity.

\begin{figure}[!t]
\centering
\begin{subfigure}{0.49\columnwidth}
\includegraphics[width=\columnwidth]{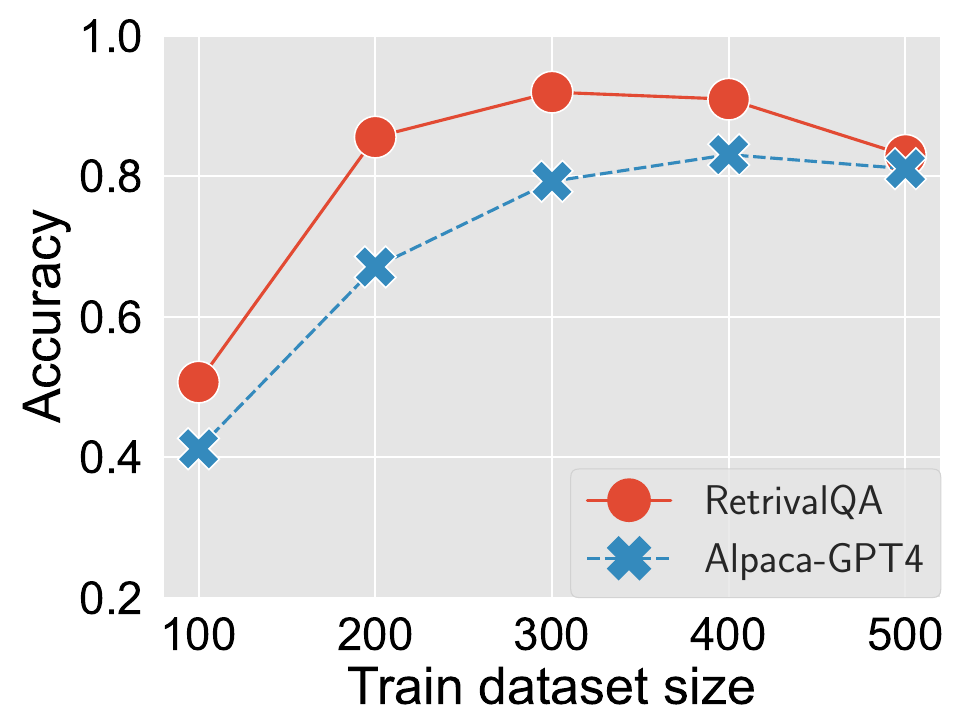}
\caption{ChatGPT}
\label{fig:size-chatgpt}
\end{subfigure}
\begin{subfigure}{0.49\columnwidth}
\includegraphics[width=\columnwidth]{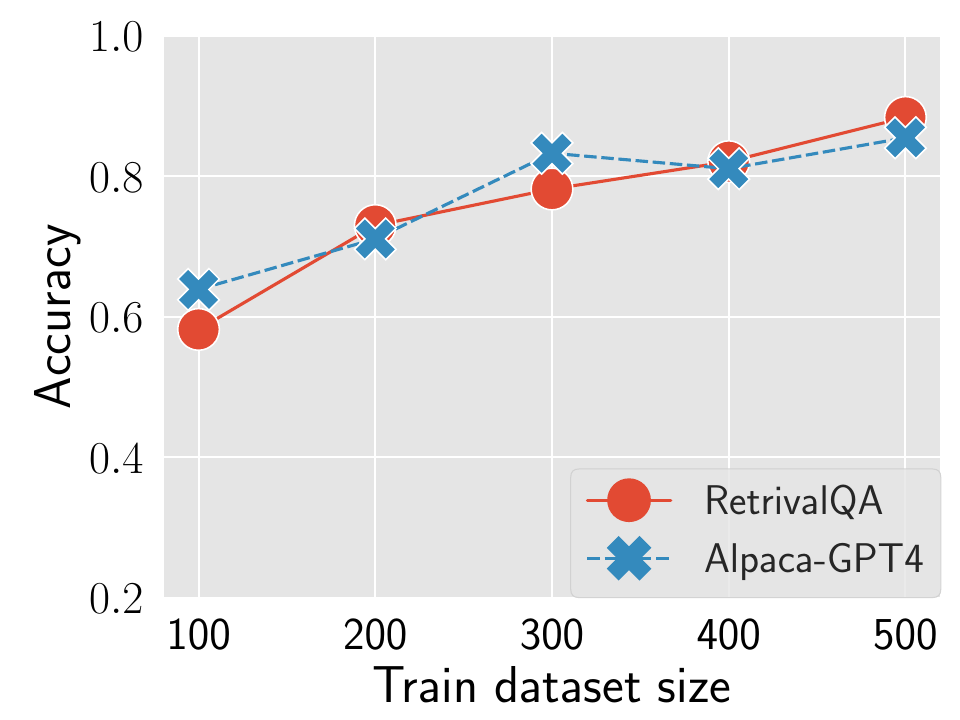}
\caption{LLaMA}
\label{fig:size-llama}
\end{subfigure}
\caption{Performance of the primary classifier of the parameter extractor with different numbers of training prompts.}
\label{fig:size-primary}
\end{figure}

\begin{figure}[!t]
\centering
\begin{subfigure}{0.49\columnwidth}
\includegraphics[width=\columnwidth]{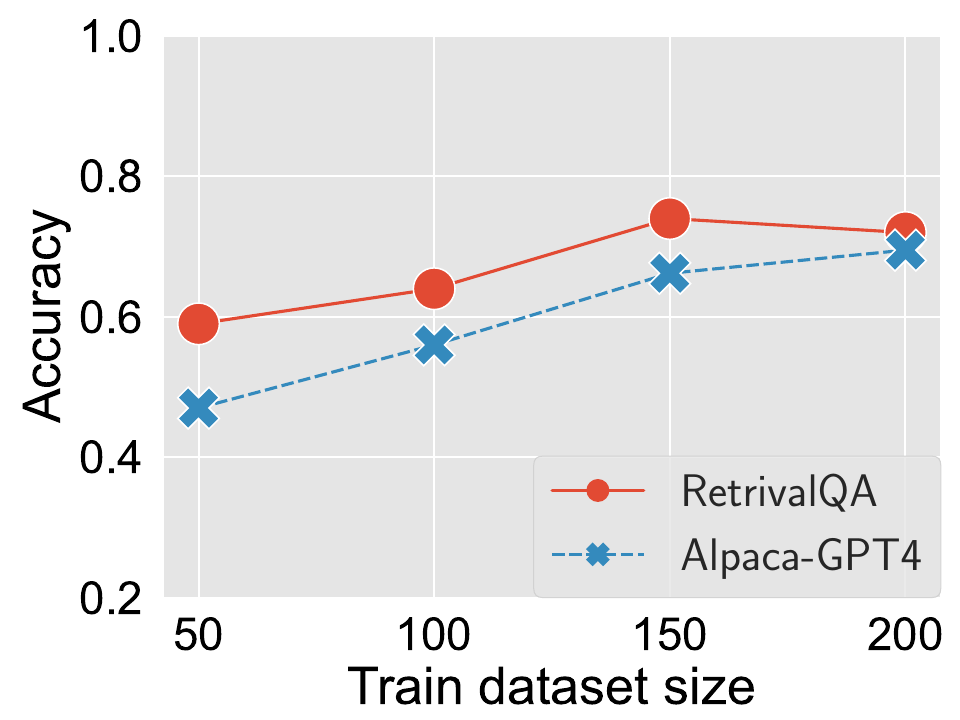}
\caption{ChatGPT}
\label{fig:size-chatgpt-role}
\end{subfigure}
\begin{subfigure}{0.49\columnwidth}
\includegraphics[width=\columnwidth]{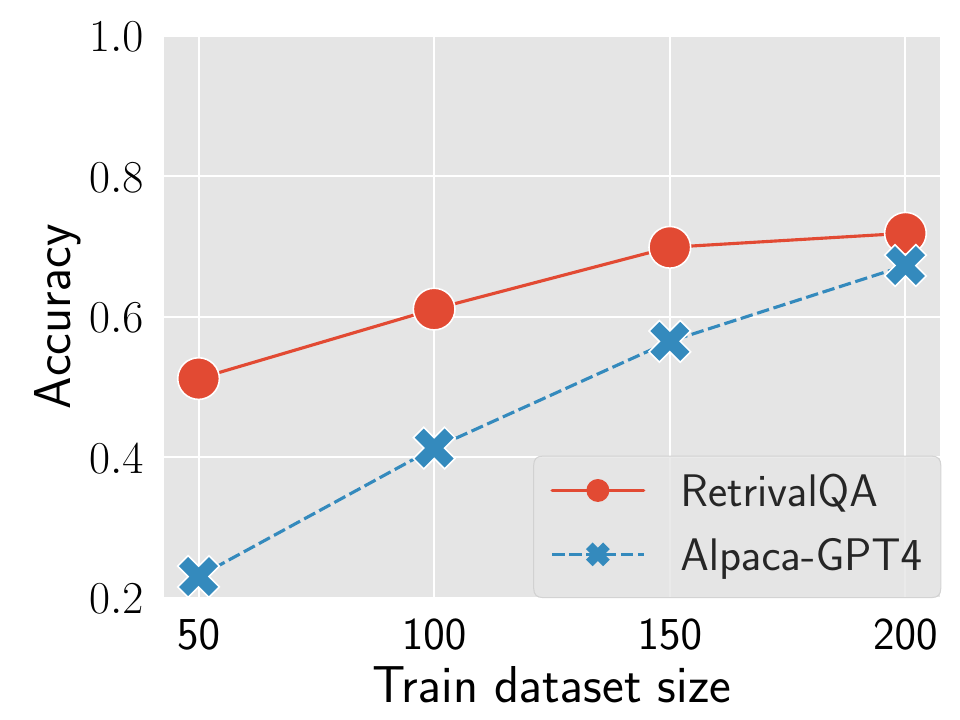}
\caption{LLaMA}
\label{fig:size-llama-role}
\end{subfigure}
\caption{Performance of the sub-classifier for role-based prompts of the parameter extractor with different numbers of training prompts.}
\label{fig:size-role}
\end{figure}

% ----------------------------------------------------
\subsection{Ablation Study}
% ----------------------------------------------------

\begin{figure}[!t]
\centering
\includegraphics[width=0.8\columnwidth]{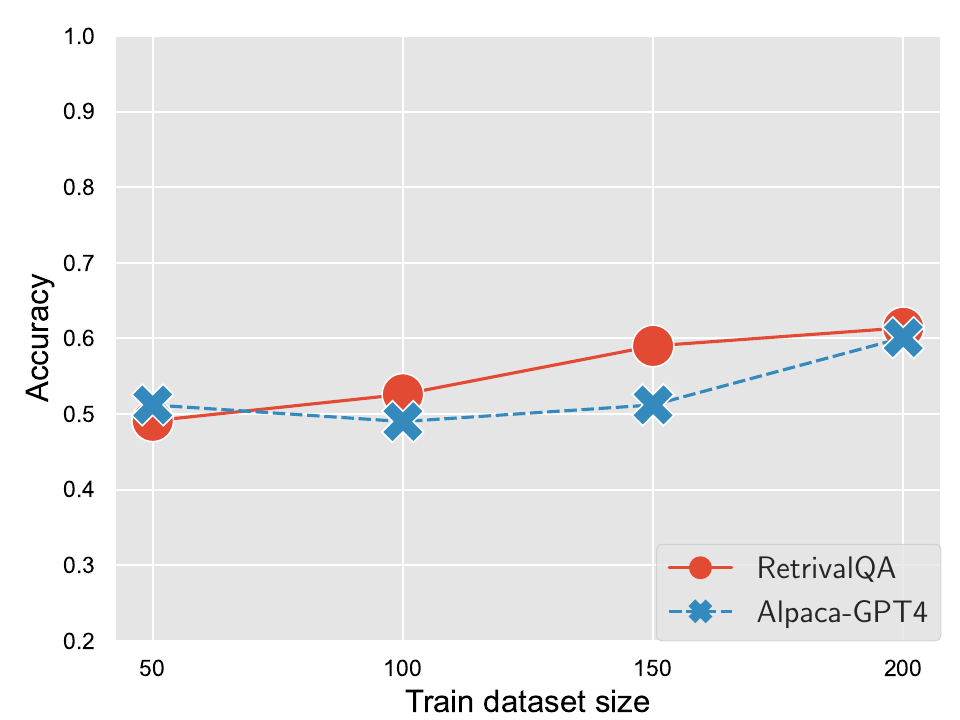}
\caption{Performance of the sub-classifier for in-context prompts of the parameter extractor with different numbers of training prompts.}
\label{fig:size-context}
\end{figure}

\mypara{Impacts of the Training Sets Size}
Due to the limited computational resources, we only leverage 500 of the prompts and their corresponding answers as the training set to train our parameter extractor.
In this part, we will analyze the impacts of the training set size on the parameter extractor's performance.
We show our results of different numbers of training prompts of the primary classifier in ~\autoref{fig:size-primary}.
We also show the results for the sub-classifier for role-based prompts in \autoref{fig:size-role} and for the sub-classifier for in-context prompts in \autoref{fig:size-context}.
Note that, as we stated before, due to the input limitation of LLaMA, we do not have the results of in-context prompts on LLaMA.
It can be seen from the figure that with more training prompts, our parameter extractor can achieve better performance.
For instance, for the primary classifier on the RetrievalQA generated by ChatGPT, with only 100 training prompts, the classifier can only achieve 0.507 accuracy, while it can be improved to 0.833 if there are 500 prompts on RetrvialQA.
It can also be observed that even the number of prompts has a great impact on the classifier's performance.
When this number comes to a certain stage, i.e., 200, the impacts will be very limited.
For instance, when the number of training prompts increases from 100 to 300, the performance of the primary classifier for Alpaca-GPT4 generated by ChatGPT can be improved from 0.412 to 0.793.
However, when the training prompts number increases from 300 to 500, the improvement in the performance is just 0.08.
Therefore, it can be concluded that although increasing the number of prompts can further improve the parameter extractors' performance, the current number of prompts is sufficient to verify the feasibility of the attack.

\begin{table}[!t]
\centering
\caption{The performance of the related roles and the randomly assigned roles.}
\label{table:abaltion-roles}
\scalebox{0.85}{
\begin{tabular}{lccc}
\toprule
Datasets & LLM & Random Role & Related Role \\
\midrule
\multirow{2}*{RetrievalQA} & ChatGPT & 0.087 & 0.732 \\
~ & LLaMA & 0.031 & 0.711 \\
\midrule
\multirow{2}*{Alpaca-GPT4} & ChatGPT & 0.073 & 0.695 \\
~ & LLaMA & 0.091 & 0.673 \\
\bottomrule
\end{tabular}
}
\end{table}

\begin{table*}[!t]
\centering
\caption{Different reconstruction prompts.
Note that the original prompt for the example is that:
Where do movies find newborn babies? 
I have always thought that maybe they are just fake or some lady really let them use her newborn.}
\label{table:reconstruct_prompt}
\scalebox{0.85}{
\begin{tabular}{p{.4\linewidth}|p{.07\linewidth}|p{.4\linewidth}}
\toprule
\textbf{Reconstruction Prompts} & \textbf{PS} & \textbf{Examples} \\
\midrule
What question you are
asked if you can generate the following answer? & 0.832 & 
How are newborn babies portrayed in movies and television shows? \\
\midrule
Assume you are the prompt reconstructor who can help the user generate the prompts based on the given answers, can you generate the original prompts for the following answer?  & 0.865 &  What methods do movies and television shows use to portray newborn babies, and what are the considerations or restrictions associated with each method? \\
\midrule
The original prompt for the answer:[Another Generated Answer] is that [Another Original Prompt]. Now what are the original prompts for the following answer? & 0.813 & Explain the techniques used in movies and television shows to portray newborn babies.\\
\bottomrule
\end{tabular}
}
\end{table*}

\mypara{Roles Assigned Based on Semantics V.S.\ Roles Assigned Randomly}
In the previous section, for the role-based prompt, we assign roles to different prompts according to the semantic correlation computed by ChatGPT.
In this section, we evaluate the performance of the sub-classifier for the role-based prompts if the roles are randomly assigned.
We show our results in \autoref{table:abaltion-roles}.
We can see from the table that when the roles are randomly assigned, the classifier cannot make the right predictions according to the answers.
For instance, on RetrievalQA, the classifier can only achieve 0.087 performance on the answers generated by ChatGPT.
Note that the poor performance of our sub-classifier on the randomly assigned roles has limited influence on the feasibility of our proposed attack.
This is because the role assigned under normal scenarios should be aligned with the following prompt.
Nevertheless, the results reveal that if the role is assigned to be aligned with the given prompt, it can boost the performance of LLMs as our classifier can easily figure out the difference between the direct prompt and the role-based prompt.
However, if the role is assigned randomly, the role-based prompt cannot improve the performance of the LLMs.

\mypara{Different Prompts for the Prompt Reconstruction}
As we have stated before, for prompt reconstruction, we leverage ChatGPT to generate the naive reversed prompt based on the answer.
However, the prompts that are used to generate the reversed prompt (we call them the reconstruction prompt) may influence the final performance.
Therefore, in this section, we study different prompts' impacts on prompt reconstruction.

As our role-based prompts and in-context prompts are generated based on the naive reversed prompts, in this section, for the sake of simplicity in the experiment, we only consider the direct prompts as the original prompts.
We collect three different reconstruction prompts and compute the similarities.
Our reconstruction prompts are designed based on the direct prompt, role-based prompt, and in-context prompt.
We show the results in \autoref{table:reconstruct_prompt}.
It can be seen that different prompts have limited influence on the reversed prompt.
For instance, the direct reconstruction prompt can achieve 0.832 similarity while the performance of the role-based reconstruction prompt is similar (0.865 similarity).
We also provide some examples of the reversed prompt in \autoref{table:reconstruct_prompt}.
It can be seen that although different reversed prompts have different structures, they all capture the key contents in the original prompts: movie and newborn baby.
Through the example, we can more intuitively see that different reconstruction prompts do not make significant impacts on the reversed prompts.

\begin{table*}[!t]
\centering
\caption{Performance of different defense methods.}
\label{table:defense}
\scalebox{0.85}{
\begin{tabular}{l|ccccccc}
\toprule
\textbf{Defense Method} & \textbf{Models} & \textbf{Datasets} & \textbf{Original PS} & \textbf{Defended PS} & \textbf{Original AS} & \textbf{Defended AS} & \textbf{Utility} \\
\midrule
\multirow{4}*{Prompt-Based Defense} & ChatGPT & RetrievalQA & 0.832 & 0.600 & 0.768 & 0.583 & 0.723 \\
~  & ChatGPT & Aplaca-GPT4 & 0.817 & 0.581 & 0.754 & 0.623 &  0.699 \\
~ & LLaMA & RetrievalQA & 0.794 & 0.551 & 0.706 & 0.512 &  0.711\\
 ~ & LLaMA & Aplaca-GPT4 & 0.712 & 0.563 & 0.712 & 0.677 &  0.745\\
 \midrule
\multirow{4}*{Answer-Based Conspicuous Defense} & ChatGPT & RetrievalQA & 0.832 & 0.664 & 0.768 & 0.486 & 0.813 \\
~  & ChatGPT & Aplaca-GPT4 & 0.817 & 0.621 & 0.754 & 0.571 & 0.832 \\
~ & LLaMA & RetrievalQA & 0.794 & 0.645 & 0.706 & 0.459 & 0.790 \\
 ~ & LLaMA & Aplaca-GPT4 & 0.712 & 0.648 & 0.712 & 0.512 & 0.811 \\
 \midrule
 \multirow{4}*{Answer-Based Inconspicuous Defense} & ChatGPT & RetrievalQA & 0.832 & 0.764 & 0.768 & 0.691 & 0.980 \\
~  & ChatGPT & Aplaca-GPT4 & 0.817 & 0.712 & 0.754 & 0.617 & 0.993 \\
~ & LLaMA & RetrievalQA & 0.794 & 0.744 & 0.706 & 0.593 & 0.978 \\
 ~ & LLaMA & Aplaca-GPT4 & 0.712 & 0.731 & 0.712 & 0.612 & 0.983 \\
\bottomrule
\end{tabular}
}
\end{table*}

% ----------------------------------------------------
\section{Defense}
% ----------------------------------------------------

As we have shown the outstanding performance of the proposed prompt stealing attacks, in this section, we discuss the potential defenses to mitigate these novel attacks.
We first summarize the defender's goal and capabilities to mitigate the prompt stealing attacks.
We propose two defense strategies, named prompt-based defense and answer-based defense, and then provide the experimental results of the proposed defense methods.

% ----------------------------------------------------
\subsection{Defender's Goals and Capabilities}
% ----------------------------------------------------

\mypara{Defender's Goals}
A defender's goal can be summarized from the following two perspectives.

\begin{itemize}
\item \mypara{Prompt Stealing Performance}
The main goal of the defender is to decrease the performance of the proposed prompt stealing attacks.
To achieve the goal, the defender is supposed to either perturb the original prompts to prevent the generated answer from leaking sensitive information, or directly perturb the generated answer.
Note that in this section, we only consider the PS and the AS of the reversed prompts and the original prompts as the metric.
This is because the final goal of the prompt stealing is to generate similar reversed prompts.
\item \mypara{Utility}
The defender is also supposed to keep the utility of the LLMs.
That means the generated answer from the defenders should be similar to the answers generated by the original prompts.
We leverage the cosine similarity to compute the similarity between the defended answer and the original answer.
The similarity here can be considered as the metric to show the utility.
Higher similarity means that the defended answer is similar to the original answer and thus can be considered to have a high utility.
\end{itemize}

\mypara{Defender's Capabilities}
The defenders are supposed to have access to the original prompts and the original generated answer.
They can perturb both the original prompts and the original answers to decrease the performance of the prompt stealing attacks.
However, normally, the defenders are also the normal users; they may not have access to the LLMs and cannot fine-tune such models.
Therefore, the deployed LLMs are supposed not to be modified to defend against the attacks.
We show later that without the modification of the LLMs, the defenders can still achieve great performance.
Also, we assume the defenders have no access to the models in the proposed attacks, which means that they cannot obtain the gradient of the classifiers in the proposed attacks.

\begin{figure*}[!t]
\centering
\includegraphics[width=2\columnwidth]{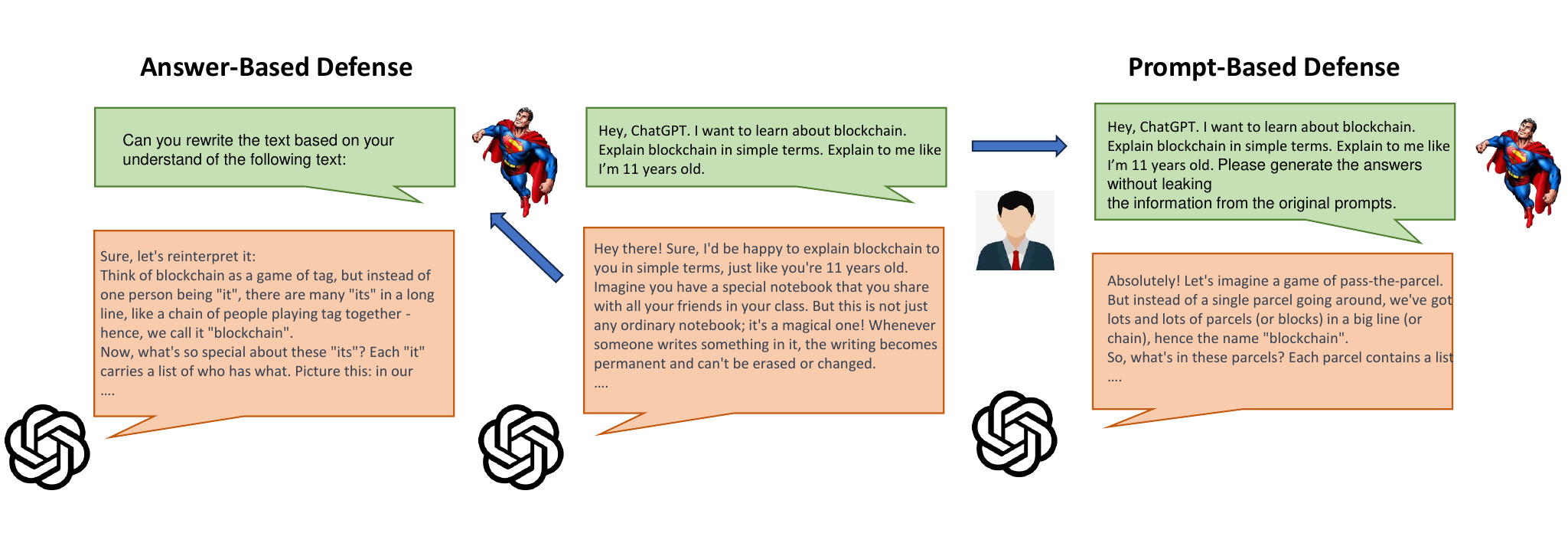}
\caption{The overview of the proposed defenses.
We consider two strategies to defend against the prompt stealing attacks, including adding perturbation on the original prompts and adding perturbation on the generated answers.}
\label{fig:defense}
\end{figure*}

% ----------------------------------------------------
\subsection{Method}
% ----------------------------------------------------

\mypara{Prompt-Based Defense}
In this defense setting, the defender aims to add some prompts to the original prompts to make the LLMs generate certain answers that contain limited information about the original prompts.
We show some examples in \autoref{fig:defense}.
The prompt we generate for the defense is predefined as ``Please generate the answers without leaking the information from the original prompts."
We hope that the added prompts can help to make the original prompts not that easy to reverse.

\mypara{Answer-Based Defense}
Another intuitive defensive method is the answer-based defense.
The answer-based defense means that the defender tries to perturb the generated answer to limit the information the adversary can obtain.
In this paper, we also leverage ChatGPT to add perturbation to the generated answers.
We consider two ways to add the perturbation with different kinds of prompts.
The first way is to ask the ChatGPT to summarize and rewrite the texts based on the generated answer.
The defense prompt can be ``Can you write the text based on your understanding of the following text: ''
We call this defense the inconspicuous defense.
Another way to add the perturbation is to make the ChatGPT understand the prompt stealing attacks and then ask it to generate similar text by removing certain critical contents.
We call this kind of defense the conspicuous defense.
The possible prompts for the conspicuous defense can be designed as ``Prompt stealing attack is an attack that aims to generate the original prompts based on the given answers; now, assume you are the defender of this attack, can you rewrite the following text to defend against that attack.''

% ----------------------------------------------------
\subsection{Results}
% ----------------------------------------------------

We show our experimental results in \autoref{table:defense}.
It can be concluded from the results that both prompt-based defense and answer-based defense can achieve good performance.
For instance, by modifying the original prompts, the AS of the attack performance can drop from 0.768 to 0.583 while keeping the high utility (0.723).
We can also conclude from the results that compared to the answer-based defense, the prompt-based defense can better mitigate the attacks while also leading to a more significant decline in the utility.
For instance, the prompt-based defense can cause the AS to drop from 0.754 to 0.623 while also causing the utility to drop to 0.699.
When comparing the two methods of answer-based defense themselves, the conspicuous defense is more effective than the inconspicuous one, but it also has a worse utility.
For instance, to defend the prompt stealing attacks on RetrievalQA prompts generated by ChatGPT, the conspicuous defense can make the AS drop from 0.768 to 0.486, while the inconspicuous defense can only cause the drop of 0.077.
However, the utility of conspicuous defense (0.813) is lower than that of the inconspicuous defense (0.980).
It is due to the fact that the inconspicuous defense does not forcibly require the ChatGPT to abandon some of the key information in the original answer, which may leak the information of the original prompts.
From the above results, we can also conclude that the defense against prompt stealing attacks is the trade-off game between the attack similarity and the utility.
If the proposed defense is very effective, then it will cause an obvious drop in the utility.
If the defense does not cause the utility to drop a lot, it tends to not be that effective in defending against the proposed attacks.
Note that all of the proposed defense methods take a lot of time to ask LLMs to make the modification for the users.
Therefore, more automated defense, such as combining all the defense prompts into the system, is needed to provide effective defense without significantly increasing the user's operational difficulty.

\mypara{Takeaway}
In conclusion, we propose three different defense strategies against the proposed attacks.
Our experimental results show that all proposed defenses can reduce the risks of the proposed attacks to a certain extent.
However, they can also cause a drop of the utility.
It can be concluded that the defense strategies we currently propose only achieve a subpar trade-off.
Therefore, we call for better defense methods.

% ----------------------------------------------------
\section{Related Work}
% ----------------------------------------------------

% ----------------------------------------------------
\subsection{Security Issues of LLMs}
% ----------------------------------------------------

Recent advancements in LLMs have facilitated a broad spectrum of applications, ranging from chatbots~\cite{DABSHBR21,RDGJWLXOSBW21,SJRDBW20,ZSGCBGGLD19,WKJK23} to the medical diagnoses~\cite{WZOWS23}.
As LLMs become more prevalent in various domains, researchers begin to study and address the potential vulnerabilities of the LLMs.
Currently, the security research regarding the LLMs develops around the following aspects: toxicity, stereotypes, adversary robustness, and privacy issues.
Previous studies.~\cite{DMRKN23,LBLTSYZNWKNYYZCMRAHZDLRRYWSOZYSKGCKHHCXSGHIZCWLMZK22,PHSCRAGMI22} have discussed the possible prompts to trigger the LLMs to generate the toxic contents.
Deshpande et al.~\cite{DMRKN23} found that with the proper roles assigned to the LLMs, it can be very mean when generating normal contents.
Liang et al.~\cite{LBLTSYZNWKNYYZCMRAHZDLRRYWSOZYSKGCKHHCXSGHIZCWLMZK22} study what types of stereotypes are more likely to be generated by the LLMs.
Regarding adversary robustness, AdvGLUE~\cite{WXWGCGAL21} was proposed to benchmark the LLMs with different state-of-the-art adversary example attacks.
There are also several papers to discuss the privacy issues, such as membership inference attacks~\cite{SSSS17,HSSDYZ21} or attribution inference attacks~\cite{MSCS19,SS20} against LLMs.
However, currently, there is no paper that views the security issues of LLMs from the perspective of the prompts.
In this paper, we propose the first prompt stealing attacks to prove that the security issues of the prompts are also a significant part of the security issues of LLMs.

\subsection{Prompt Stealing Attacks}
Before the advent of the current work, a substantial amount of research was dedicated to conducting prompt stealing attacks against text-to-image generation models~\cite{RBLEO22,RDNCC22}.
The simplest way~\cite{LLXH22,MHB21} is to leverage the image captioning models.
In particular, a remarkable proposal by Mokady et al.~\cite{MHB21} introduced ClipCap, a strategy that capitalizes on CLIP~\cite{RKHRGASAMCKS21} to translate an image into a prompt effectively.

Taking this concept further, Shen et al.~\cite{SQBZ23} went on to propose specialized datasets that could be employed to carry out prompt stealing attacks against text-to-image models.
These datasets provided a novel resource for researchers, opening up new opportunities for exploring the vulnerabilities of text-to-image models and refining the techniques used for prompt stealing attacks.

With respect to LLMs, Zhang et al.~\cite{ZI23} made the intriguing discovery that the system prompt can be readily reversed.
This finding implies that LLMs, while robust and powerful, may still be vulnerable to prompt stealing attacks, potentially presenting a significant concern in LLM security.
However, there are no prompt stealing attacks based on the generated answers from LLMs currently.

% ----------------------------------------------------
\section{Conclusion}
% ----------------------------------------------------

In this paper, we propose the first prompt stealing attacks against LLMs.
Our prompt stealing attacks aim to steal the prompts based on the corresponding answers generated by the LLMs.
Our attacks contain two hierarchical parts: the parameter extractor and the prompt reconstructor.
The goal of the parameter extractor is to predict the properties of the original prompts.
We first observe that most of the prompts fall into three categories: direct prompt, role-based prompt, and in-context prompt.
Therefore, the first classifier in the proposed parameter extractor is to predict which kind of prompts the original prompts are.
Based on that prediction, two sub-classifiers are developed to determine the further role or the number of contexts.
Following the parameter extractor, the prompt reconstructor is developed to generate the reversed prompts, which are assumed to be similar to the target prompts.
We leverage ChatGPT as the backbone of our prompt reconstructor to generate the naive reversed prompts based on the given answers.
Then, the previous information is used to modify the naive reversed prompts.
To be more specific, if the parameter extractor predicts the original prompts to be role-based prompts, then we will add the role-based prompt to the naive prompts to construct the new reversed prompts.
If the original prompts are predicted to be the in-context prompts, then a certain number of the contexts will be generated and added to the naive prompts as the reversed prompts.
Extensive results show that the proposed prompt stealing attacks can have great performance.
In particular, the parameter can successfully predict the type of the original prompts and further information.
Armed with the parameter information, the prompt reconstructor can generate more similar reversed prompts.
We also propose two possible defenses against prompt stealing attacks, and the results show there is a fair trade-off between the defense performance and the utility.

Our research emphasizes that prompts are not secrets and can be easily stolen from their generated responses, leading us to urge the scientific community to focus on the exploration and design of more effective defenses against prompt stealing attacks.
We also look forward to more security research and discussions on LLMs from the perspective of the prompt.

% ----------------------------------------------------
\newpage
\begin{small}
\bibliographystyle{plain}
\bibliography{normal_generated_py3}

\begin{thebibliography}{10}

\bibitem{chatgpt}
\url{https://chat.openai.com/chat}.

\bibitem{AAKV23}
Arian Askari, Mohammad Aliannejadi, Evangelos Kanoulas, and Suzan Verberne.
\newblock {Generating Synthetic Documents for Cross-Encoder Re-Rankers: {A} Comparative Study of ChatGPT and Human Experts}.
\newblock {\em {CoRR abs/2305.02320}}, 2023.

\bibitem{BMRSKDNSSAAHKHCRZWWHCSLGCCBMRSA20}
Tom~B. Brown, Benjamin Mann, Nick Ryder, Melanie Subbiah, Jared Kaplan, Prafulla Dhariwal, Arvind Neelakantan, Pranav Shyam, Girish Sastry, Amanda Askell, Sandhini Agarwal, Ariel Herbert{-}Voss, Gretchen Krueger, Tom Henighan, Rewon Child, Aditya Ramesh, Daniel~M. Ziegler, Jeffrey Wu, Clemens Winter, Christopher Hesse, Mark Chen, Eric Sigler, Mateusz Litwin, Scott Gray, Benjamin Chess, Jack Clark, Christopher Berner, Sam McCandlish, Alec Radford, Ilya Sutskever, and Dario Amodei.
\newblock {Language Models are Few-Shot Learners}.
\newblock In {\em {Annual Conference on Neural Information Processing Systems (NeurIPS)}}. NeurIPS, 2020.

\bibitem{CMWWLGWYSXTD23}
Yuzhe Cai, Shaoguang Mao, Wenshan Wu, Zehua Wang, Yaobo Liang, Tao Ge, Chenfei Wu, Wang You, Ting Song, Yan Xia, Jonathan Tien, and Nan Duan.
\newblock {Low-code {LLM:} Visual Programming over LLMs}.
\newblock {\em {CoRR abs/2304.08103}}, 2023.

\bibitem{CLBMLA17}
Paul~F. Christiano, Jan Leike, Tom~B. Brown, Miljan Martic, Shane Legg, and Dario Amodei.
\newblock {Deep Reinforcement Learning from Human Preferences}.
\newblock In {\em {Annual Conference on Neural Information Processing Systems (NIPS)}}, pages 4299--4307. NIPS, 2017.

\bibitem{DLLWZLWZL23}
Gelei Deng, Yi~Liu, Yuekang Li, Kailong Wang, Ying Zhang, Zefeng Li, Haoyu Wang, Tianwei Zhang, and Yang Liu.
\newblock {Jailbreaker: Automated Jailbreak Across Multiple Large Language Model Chatbots}.
\newblock {\em {CoRR abs/2307.08715}}, 2023.

\bibitem{DMRKN23}
Ameet Deshpande, Vishvak Murahari, Tanmay Rajpurohit, Ashwin Kalyan, and Karthik Narasimhan.
\newblock {Toxicity in ChatGPT: Analyzing Persona-assigned Language Models}.
\newblock {\em {CoRR abs/2304.05335}}, 2023.

\bibitem{DCLT19}
Jacob Devlin, Ming{-}Wei Chang, Kenton Lee, and Kristina Toutanova.
\newblock {BERT: Pre-training of Deep Bidirectional Transformers for Language Understanding}.
\newblock In {\em {Conference of the North American Chapter of the Association for Computational Linguistics: Human Language Technologies (NAACL-HLT)}}, pages 4171--4186. ACL, 2019.

\bibitem{DABSHBR21}
Emily Dinan, Gavin Abercrombie, A.~Stevie Bergman, Shannon~L. Spruit, Dirk Hovy, Y{-}Lan Boureau, and Verena Rieser.
\newblock {Anticipating Safety Issues in {E2E} Conversational {AI:} Framework and Tooling}.
\newblock {\em {CoRR abs/2107.03451}}, 2021.

\bibitem{DFWUKW20}
Emily Dinan, Angela Fan, Adina Williams, Jack Urbanek, Douwe Kiela, and Jason Weston.
\newblock {Queens are Powerful too: Mitigating Gender Bias in Dialogue Generation}.
\newblock In {\em {Conference on Empirical Methods in Natural Language Processing (EMNLP)}}, pages 8173--8188. ACL, 2020.

\bibitem{GAG23}
Eduardo~C. Garrido{-}Merch{\'{a}}n, Jos{\'{e}}~Luis Arroyo{-}Barrig{\"{u}}ete, and Roberto Gozalo{-}Brizuela.
\newblock {Simulating {H.P.} Lovecraft horror literature with the ChatGPT large language model}.
\newblock {\em {CoRR abs/2305.03429}}, 2023.

\bibitem{GZWJNDYW23}
Biyang Guo, Xin Zhang, Ziyuan Wang, Minqi Jiang, Jinran Nie, Yuxuan Ding, Jianwei Yue, and Yupeng Wu.
\newblock {How Close is ChatGPT to Human Experts? Comparison Corpus, Evaluation, and Detection}.
\newblock {\em {CoRR abs/2301.07597}}, 2023.

\bibitem{HSCBZ23}
Xinlei He, Xinyue Shen, Zeyuan Chen, Michael Backes, and Yang Zhang.
\newblock {MGTBench: Benchmarking Machine-Generated Text Detection}.
\newblock {\em {CoRR abs/2303.14822}}, 2023.

\bibitem{HSSDYZ21}
Hongsheng Hu, Zoran Salcic, Lichao Sun, Gillian Dobbie, Philip~S. Yu, and Xuyun Zhang.
\newblock {Membership Inference Attacks on Machine Learning: A Survey}.
\newblock {\em {ACM Computing Surveys}}, 2021.

\bibitem{KLSGZH23}
Daniel Kang, Xuechen Li, Ion Stoica, Carlos Guestrin, Matei Zaharia, and Tatsunori Hashimoto.
\newblock {Exploiting Programmatic Behavior of LLMs: Dual-Use Through Standard Security Attacks}.
\newblock {\em {CoRR abs/2302.05733}}, 2023.

\bibitem{LGFXS23}
Haoran Li, Dadi Guo, Wei Fan, Mingshi Xu, and Yangqiu Song.
\newblock {Multi-step Jailbreaking Privacy Attacks on ChatGPT}.
\newblock {\em {CoRR abs/2304.05197}}, 2023.

\bibitem{LLXH22}
Junnan Li, Dongxu Li, Caiming Xiong, and Steven C.~H. Hoi.
\newblock {{BLIP:} Bootstrapping Language-Image Pre-training for Unified Vision-Language Understanding and Generation}.
\newblock {\em {CoRR abs/2201.12086}}, 2022.

\bibitem{LBLTSYZNWKNYYZCMRAHZDLRRYWSOZYSKGCKHHCXSGHIZCWLMZK22}
Percy Liang, Rishi Bommasani, Tony Lee, Dimitris Tsipras, Dilara Soylu, Michihiro Yasunaga, Yian Zhang, Deepak Narayanan, Yuhuai Wu, Ananya Kumar, Benjamin Newman, Binhang Yuan, Bobby Yan, Ce~Zhang, Christian Cosgrove, Christopher~D. Manning, Christopher R{\'{e}}, Diana Acosta{-}Navas, Drew~A. Hudson, Eric Zelikman, Esin Durmus, Faisal Ladhak, Frieda Rong, Hongyu Ren, Huaxiu Yao, Jue Wang, Keshav Santhanam, Laurel~J. Orr, Lucia Zheng, Mert Y{\"{u}}ksekg{\"{o}}n{\"{u}}l, Mirac Suzgun, Nathan Kim, Neel Guha, Niladri~S. Chatterji, Omar Khattab, Peter Henderson, Qian Huang, Ryan Chi, Sang~Michael Xie, Shibani Santurkar, Surya Ganguli, Tatsunori Hashimoto, Thomas Icard, Tianyi Zhang, Vishrav Chaudhary, William Wang, Xuechen Li, Yifan Mai, Yuhui Zhang, and Yuta Koreeda.
\newblock {Holistic Evaluation of Language Models}.
\newblock {\em {CoRR abs/2211.09110}}, 2022.

\bibitem{LDXLZZZZL23}
Yi~Liu, Gelei Deng, Zhengzi Xu, Yuekang Li, Yaowen Zheng, Ying Zhang, Lida Zhao, Tianwei Zhang, and Yang Liu.
\newblock {Jailbreaking ChatGPT via Prompt Engineering: An Empirical Study}.
\newblock {\em {CoRR abs/2305.13860}}, 2023.

\bibitem{MSCS19}
Luca Melis, Congzheng Song, Emiliano~De Cristofaro, and Vitaly Shmatikov.
\newblock {Exploiting Unintended Feature Leakage in Collaborative Learning}.
\newblock In {\em {IEEE Symposium on Security and Privacy (S\&P)}}, pages 497--512. IEEE, 2019.

\bibitem{MLKMF23}
Eric Mitchell, Yoonho Lee, Alexander Khazatsky, Christopher~D. Manning, and Chelsea Finn.
\newblock {DetectGPT: Zero-Shot Machine-Generated Text Detection using Probability Curvature}.
\newblock {\em {CoRR abs/2301.11305}}, 2023.

\bibitem{MHB21}
Ron Mokady, Amir Hertz, and Amit~H. Bermano.
\newblock {ClipCap: {CLIP} Prefix for Image Captioning}.
\newblock {\em {CoRR abs/2111.09734}}, 2021.

\bibitem{O23}
OpenAI.
\newblock {{GPT-4} Technical Report}.
\newblock {\em {CoRR abs/2303.08774}}, 2023.

\bibitem{POCMLB23}
Joon~Sung Park, Joseph~C. O'Brien, Carrie~J. Cai, Meredith~Ringel Morris, Percy Liang, and Michael~S. Bernstein.
\newblock {Generative Agents: Interactive Simulacra of Human Behavior}.
\newblock {\em {CoRR abs/2304.03442}}, 2023.

\bibitem{PLHGG23}
Baolin Peng, Chunyuan Li, Pengcheng He, Michel Galley, and Jianfeng Gao.
\newblock {Instruction Tuning with {GPT-4}}.
\newblock {\em {CoRR abs/2304.03277}}, 2023.

\bibitem{PHSCRAGMI22}
Ethan Perez, Saffron Huang, H.~Francis Song, Trevor Cai, Roman Ring, John Aslanides, Amelia Glaese, Nat McAleese, and Geoffrey Irving.
\newblock {Red Teaming Language Models with Language Models}.
\newblock {\em {CoRR abs/2202.03286}}, 2022.

\bibitem{RKHRGASAMCKS21}
Alec Radford, Jong~Wook Kim, Chris Hallacy, Aditya Ramesh, Gabriel Goh, Sandhini Agarwal, Girish Sastry, Amanda Askell, Pamela Mishkin, Jack Clark, Gretchen Krueger, and Ilya Sutskever.
\newblock {Learning Transferable Visual Models From Natural Language Supervision}.
\newblock In {\em {International Conference on Machine Learning (ICML)}}, pages 8748--8763. PMLR, 2021.

\bibitem{AKTI20}
Alec Radford, Karthik Narasimhan, Tim Salimans, and Ilya Sutskever.
\newblock Improving language understanding by generative pre-training.
\newblock 2018.

\bibitem{RWCLAS19}
Alec Radford, Jeffrey Wu, Rewon Child, David Luan, Dario Amodei, and Ilya Sutskever.
\newblock {Language Models are Unsupervised Multitask Learners}.
\newblock {\em {OpenAI blog}}, 2019.

\bibitem{RDNCC22}
Aditya Ramesh, Prafulla Dhariwal, Alex Nichol, Casey Chu, and Mark Chen.
\newblock {Hierarchical Text-Conditional Image Generation with CLIP Latents}.
\newblock {\em {CoRR abs/2204.06125}}, 2022.

\bibitem{RG19}
Nils Reimers and Iryna Gurevych.
\newblock {Sentence-BERT: Sentence Embeddings using Siamese BERT-Networks}.
\newblock In {\em {Conference on Empirical Methods in Natural Language Processing and International Joint Conference on Natural Language Processing (EMNLP-IJCNLP)}}, pages 3980--3990. ACL, 2019.

\bibitem{RDGJWLXOSBW21}
Stephen Roller, Emily Dinan, Naman Goyal, Da~Ju, Mary Williamson, Yinhan Liu, Jing Xu, Myle Ott, Eric~Michael Smith, Y{-}Lan Boureau, and Jason Weston.
\newblock {Recipes for Building an Open-Domain Chatbot}.
\newblock In {\em {Conference of the European Chapter of the Association for Computational Linguistics (EACL)}}, pages 300--325. ACL, 2021.

\bibitem{RBLEO22}
Robin Rombach, Andreas Blattmann, Dominik Lorenz, Patrick Esser, and Bj{\"{o}}rn Ommer.
\newblock {High-Resolution Image Synthesis with Latent Diffusion Models}.
\newblock In {\em {IEEE Conference on Computer Vision and Pattern Recognition (CVPR)}}, pages 10684--10695. IEEE, 2022.

\bibitem{SCBZ23}
Xinyue Shen, Zeyuan Chen, Michael Backes, and Yang Zhang.
\newblock {In ChatGPT We Trust? Measuring and Characterizing the Reliability of ChatGPT}.
\newblock {\em {CoRR abs/2304.08979}}, 2023.

\bibitem{SQBZ23}
Xinyue Shen, Yiting Qu, Michael Backes, and Yang Zhang.
\newblock {Prompt Stealing Attacks Against Text-to-Image Generation Models}.
\newblock {\em {CoRR abs/2302.09923}}, 2023.

\bibitem{SSSS17}
Reza Shokri, Marco Stronati, Congzheng Song, and Vitaly Shmatikov.
\newblock {Membership Inference Attacks Against Machine Learning Models}.
\newblock In {\em {IEEE Symposium on Security and Privacy (S\&P)}}, pages 3--18. IEEE, 2017.

\bibitem{SJRDBW20}
Kurt Shuster, Da~Ju, Stephen Roller, Emily Dinan, Y{-}Lan Boureau, and Jason Weston.
\newblock {The Dialogue Dodecathlon: Open-Domain Knowledge and Image Grounded Conversational Agents}.
\newblock In {\em {Annual Meeting of the Association for Computational Linguistics (ACL)}}, pages 2453--2470. ACL, 2020.

\bibitem{SBBCSZZ22}
Wai~Man Si, Michael Backes, Jeremy Blackburn, Emiliano~De Cristofaro, Gianluca Stringhini, Savvas Zannettou, and Yang Zhang.
\newblock {Why So Toxic? Measuring and Triggering Toxic Behavior in Open-Domain Chatbots}.
\newblock In {\em {ACM SIGSAC Conference on Computer and Communications Security (CCS)}}, pages 2659--2673. ACM, 2022.

\bibitem{SBHP23}
Dominik Sobania, Martin Briesch, Carol Hanna, and Justyna Petke.
\newblock {An Analysis of the Automatic Bug Fixing Performance of ChatGPT}.
\newblock {\em {CoRR abs/2301.08653}}, 2023.

\bibitem{SS20}
Congzheng Song and Vitaly Shmatikov.
\newblock {Overlearning Reveals Sensitive Attributes}.
\newblock In {\em {International Conference on Learning Representations (ICLR)}}, 2020.

\bibitem{SOWZLVRAC20}
Nisan Stiennon, Long Ouyang, Jeff Wu, Daniel~M. Ziegler, Ryan Lowe, Chelsea Voss, Alec Radford, Dario Amodei, and Paul~F. Christiano.
\newblock {Learning to summarize from human feedback}.
\newblock {\em {CoRR abs/2009.01325}}, 2020.

\bibitem{SWG23}
Hengky Susanto, David~James Woo, and Kai Guo.
\newblock {The Role of {AI} in Human-AI Creative Writing for Hong Kong Secondary Students}.
\newblock {\em {CoRR abs/2304.11276}}, 2023.

\bibitem{TLIMLLRGHARJGL23}
Hugo Touvron, Thibaut Lavril, Gautier Izacard, Xavier Martinet, Marie{-}Anne Lachaux, Timoth{\'{e}}e Lacroix, Baptiste Rozi{\`{e}}re, Naman Goyal, Eric Hambro, Faisal Azhar, Aur{\'{e}}lien Rodriguez, Armand Joulin, Edouard Grave, and Guillaume Lample.
\newblock {LLaMA: Open and Efficient Foundation Language Models}.
\newblock {\em {CoRR abs/2302.13971}}, 2023.

\bibitem{VSPUJGKP17}
Ashish Vaswani, Noam Shazeer, Niki Parmar, Jakob Uszkoreit, Llion Jones, Aidan~N. Gomez, Lukasz Kaiser, and Illia Polosukhin.
\newblock {Attention is All you Need}.
\newblock In {\em {Annual Conference on Neural Information Processing Systems (NIPS)}}, pages 5998--6008. NIPS, 2017.

\bibitem{WXWGCGAL21}
Boxin Wang, Chejian Xu, Shuohang Wang, Zhe Gan, Yu~Cheng, Jianfeng Gao, Ahmed~Hassan Awadallah, and Bo~Li.
\newblock {Adversarial {GLUE:} {A} Multi-Task Benchmark for Robustness Evaluation of Language Models}.
\newblock In {\em {Annual Conference on Neural Information Processing Systems (NeurIPS)}}. NeurIPS, 2021.

\bibitem{WZOWS23}
Sheng Wang, Zihao Zhao, Xi~Ouyang, Qian Wang, and Dinggang Shen.
\newblock {ChatCAD: Interactive Computer-Aided Diagnosis on Medical Image using Large Language Models}.
\newblock {\em {CoRR abs/2302.07257}}, 2023.

\bibitem{WWSLCNCZ23}
Xuezhi Wang, Jason Wei, Dale Schuurmans, Quoc~V. Le, Ed~H. Chi, Sharan Narang, Aakanksha Chowdhery, and Denny Zhou.
\newblock {Self-Consistency Improves Chain of Thought Reasoning in Language Models}.
\newblock In {\em {International Conference on Learning Representations (ICLR)}}, 2023.

\bibitem{WWSBIXCLZ22}
Jason Wei, Xuezhi Wang, Dale Schuurmans, Maarten Bosma, Brian Ichter, Fei Xia, Ed~H. Chi, Quoc~V. Le, and Denny Zhou.
\newblock {Chain-of-Thought Prompting Elicits Reasoning in Large Language Models}.
\newblock In {\em {Annual Conference on Neural Information Processing Systems (NeurIPS)}}. NeurIPS, 2022.

\bibitem{WKJK23}
Jing Wei, Sungdong Kim, Hyunhoon Jung, and Young{-}Ho Kim.
\newblock {Leveraging Large Language Models to Power Chatbots for Collecting User Self-Reported Data}.
\newblock {\em {CoRR abs/2301.05843}}, 2023.

\bibitem{YYZSGCN23}
Shunyu Yao, Dian Yu, Jeffrey Zhao, Izhak Shafran, Thomas~L. Griffiths, Yuan Cao, and Karthik Narasimhan.
\newblock {Tree of Thoughts: Deliberate Problem Solving with Large Language Models}.
\newblock {\em {CoRR abs/2305.10601}}, 2023.

\bibitem{ZWMG22}
Eric Zelikman, Yuhuai Wu, Jesse Mu, and Noah~D. Goodman.
\newblock {STaR: Bootstrapping Reasoning With Reasoning}.
\newblock In {\em {Annual Conference on Neural Information Processing Systems (NeurIPS)}}. NeurIPS, 2022.

\bibitem{ZYJ23}
Xiaoli Zhang, Daksha Yadav, and Boyang~Tom Jin.
\newblock {Cash Transaction Booking via Retrieval Augmented LLM}.
\newblock In {\em {ACM Conference on Knowledge Discovery and Data Mining (KDD)}}. ACM, 2023.

\bibitem{ZI23}
Yiming Zhang and Daphne Ippolito.
\newblock {Prompts Should not be Seen as Secrets: Systematically Measuring Prompt Extraction Attack Success}.
\newblock {\em {CoRR abs/2307.06865}}, 2023.

\bibitem{ZSGCBGGLD19}
Yizhe Zhang, Siqi Sun, Michel Galley, Yen{-}Chun Chen, Chris Brockett, Xiang Gao, Jianfeng Gao, Jingjing Liu, and Bill Dolan.
\newblock {DialoGPT: Large-Scale Generative Pre-training for Conversational Response Generation}.
\newblock {\em {CoRR abs/1911.00536}}, 2019.

\end{thebibliography}
\end{small}
% ----------------------------------------------------

% ----------------------------------------------------
\appendix
% ----------------------------------------------------

\begin{table*}[!t]
\centering
\caption{Commonly used roles.}
\label{table:commonroles}
\scalebox{0.85}{
\begin{tabular}{l| l}
\toprule
\textbf{Prompt Role} & \textbf{Description} \\
\midrule
Science Information Provider &  LLMs are assumed to provide information related to scientific topics and research \\
\midrule
Media Information Provider  &  LLMs are assumed to offer information about news, events, and other entertainment activities  \\
\midrule
Ethical/Sex Information Provider &  LLMs are assumed to provide information pertaining to ethics or sexual topics  \\
\midrule
Legal Advisor &  LLMs are assumed to provide legal matters information \\
\midrule
Research Analyst &  LLMs are assumed to help the researcher analyze and present experiment results \\
\midrule
Movie/Artwork Advisor & LLMs are assumed to recommend movies or other artworks based on the preference \\
\midrule
Education Information Provider & LLMs are assumed to provide information or resources on the education-related topic \\
\midrule
Political Information Provider & LLMs are assumed to help the people analyze information about political events and issues \\
\midrule
Customer Support & LLMs are assumed to assist customers with inquiries and support-related matters \\
\midrule
Translator & LLMs are assumed to translate one language into another \\
\midrule
Financial Advisor &  LLMs are assumed to offer financial advice and guidance.\\
\midrule
Health Information Provider &  LLMs are assumed to provide information on health-related topics \\
\midrule
History/Cultural Information Provider &  LLMs are assumed to offer information about historical and cultural subjects \\
\midrule
Fact Checker & LLMs are assumed to verify and confirm the accuracy of information  \\
\midrule
Sports Analyst & LLMs are assumed to help people understand sport-related events and data \\
\bottomrule
\end{tabular}
}
\end{table*}

% ----------------------------------------------------
\end{document}